\documentclass{article}

\usepackage[preprint]{neurips_2026}

\usepackage[utf8]{inputenc}
\usepackage[T1]{fontenc}
\usepackage{hyperref}
\usepackage{url}
\usepackage{booktabs}
\usepackage{amsmath,amssymb,amsfonts}
\usepackage{nicefrac}
\usepackage{microtype}
\usepackage{xcolor}
\usepackage{graphicx}
\usepackage{multirow}
\usepackage{enumitem}
\usepackage{tikz}
\usetikzlibrary{shapes.geometric, arrows.meta, positioning}
\usepackage{float}
\usepackage[compact]{titlesec}
\usepackage{caption}
\usepackage{sidecap}
\sidecaptionvpos{figure}{t}

\newcommand{\fb}{\textsc{FormBench}}
\newcommand{\ndcg}{nDCG@10}

\title{\fb{}: A Benchmark for Fine-Grained Retrieval in Formulation Patents}

\author{%
  Vineeth Venugopal \\
  Reciprocal Space, India \\
  \texttt{vinven7@gmail.com} \\
  \And
  Elsa Olivetti \\
  MIT, USA \\
  \And
  Jonathan Welch \\
  Albert Invent, USA \\
  \texttt{jonathan@albertinvent.com} \\
}

\begin{document}

\maketitle

\begin{abstract}
We introduce \fb{}, a platform for
testing the ability of models to retrieve precise chemical information from
technical documents. \fb{} contains 55,352 queries and two retrieval corpus
variants each with $\sim$1M passages: \fb{}-Random (C0), with randomly sampled distractors, and
\fb{}-Structured (C1), with structurally proximate distractors. Each query is
paired with an anchor passage, a hard negative, and a soft negative ---
derived from a materials ontology through entity overlap scoring, providing
a deeper graded signal on a model's discriminatory power. We evaluate 10
model configurations spanning commercial API encoders (OpenAI, Voyage),
open-source encoders (Nomic, GTE-large, Qwen3-Embed), and their
task-adaptive fine-tuned (TAPT) variants. We find: (1) commercial APIs
score very poorly across all configurations (OpenAI-3-large: 0.009 on C1,
0.213 on C0); (2) TAPT on \fb{} triplets lifts Qwen3-Embed to 0.4085 on C1, 0.4835 on C0;
(3) BM25 remains competitive (0.375 on C1, 0.467 on C0), exceeded only by
TAPT-Qwen3; (4) corpus size has a significant impact on retrieval
behavior, as demonstrated through a corpus-size ablation; (5) the API
failure traces to within-domain geometric collapse; (6) APIs outperform
even the TAPT variants on classification and clustering tasks. \fb{} is
released in BEIR-compatible format
under CC-BY-4.0.
\end{abstract}

\section{Introduction}
\label{sec:intro}

Domain-specific scientific information retrieval from unstructured text is
harder than general retrieval. Relevant passages may share the same tokens
representing domain-specific entities --- surfactants, polymers, solvents,
reaction conditions, performance descriptors, etc --- yet differ in the
functional roles those entities play and the application context for which
a system was designed. The discriminative signal is ontological, not
topical: the same surfactant may act as emulsifier, dispersant, or wetting
agent depending on context, and it often takes technical training to
discern the difference. Language models trained on generic web data are
not designed for this, and consequently existing text retrieval benchmarks
do not expose this failure mode.

We introduce \fb{} (\textbf{Form}ulation \textbf{Bench}mark), a
large-scale retrieval benchmark of 55,352 queries built from US
formulation patents. Distractors are functional near-misses --- passages
sharing high chemical entity density with the anchor but differing in
role, context, and function --- not topic distractors. Two corpus variants
isolate distractor difficulty: \textbf{FormBench-Random} (C0) ($\sim$997K
random-distractor passages) tests coarse domain retrieval;
\textbf{FormBench-Structured} (C1) ($\sim$994K ontologically adjacent near-miss
passages) requires functional-level discrimination. Both share identical
queries and relevance judgments and are released in BEIR format.

We evaluate 10 model configurations spanning commercial API encoders
(OpenAI, Voyage), open-source encoders (Nomic, GTE-large, Qwen3-Embed),
and their task-adaptive fine-tuned (TAPT) variants, along with a BM25
lexical baseline. Three findings emerge. First, commercial APIs score
below the BM25 lexical floor (\ndcg{} $\leq$ 0.009
vs.\ 0.375). Second, TAPT on taxonomy-derived triplets performs significantly
better than the API baselines and exceeds the
lexical ceiling (Qwen3-Embed + MNRL: 0.4085 on C1, 0.4835 on C0). Third, the embedding
modification produced by TAPT is task-specific: models that excel at
retrieval gain nothing on classification or clustering, yet they retain
their advantage under corpus-size ablation. Together these findings point
to a single mechanism: general embeddings struggle with ultra-fine-grained
discrimination between structurally similar technical passages unless
trained explicitly for that retrieval geometry.

\paragraph{Contributions.}
\begin{enumerate}[leftmargin=*,itemsep=1pt]
  \item \textbf{\fb{} dataset.} 55,352 queries with ontologically derived
  graded (0--3) relevance, two purpose-designed corpus variants of
  $\sim$1M passages each, BEIR-compatible.
  \item \textbf{Empirical baselines.} 10 model configurations under base
  and TAPT conditions, plus a BM25 lexical baseline; binary and graded
  \ndcg{}, MRR, R@100; corpus-size ablation and latency analysis.
  \item \textbf{Mechanistic analysis.} Geometric characterisation of API
  model failure and evidence that TAPT on structured negatives resolves
  it --- a diagnostic invisible without within-domain distractor design.
\end{enumerate}

\section{Related Work}
\label{sec:related}

BEIR~\citep{thakur2021beir} and MTEB~\citep{muennighoff2023mteb} established zero-shot retrieval evaluation as a standard, but their predominantly general-domain corpora mask domain-specific failures. FinMTEB~\citep{tang2024finmteb}, ChemTEB~\citep{shiraee2024chemteb}, MLEB~\citep{butler2025mleb}, and SciRepEval~\citep{singh2023scirepeval} each demonstrate that high aggregate MTEB scores do not predict performance on specialised professional text, motivating purpose-built benchmarks. \fb{} follows this line for formulation chemistry; we focus on dense bi-encoder evaluation~\citep{reimers2019sbert,karpukhin2020dpr}, where embedding geometry is directly interpretable and domain adaptation is tractable.

The closest prior work is in patents and chemistry. PatentSBERTa~\citep{bekamiri2021patentsberta} and PaECTER~\citep{ghosh2024paecter} adapt sentence transformers to patent similarity via contrastive and citation-informed training. PatenTEB~\citep{ayaou2025patenteb} provides 15 benchmark tasks across 109 IPC subclasses; DAPFAM~\citep{ayaou2025dapfam} addresses patent-family prior-art retrieval. Both use binary, citation-derived relevance over title and abstract --- neither targets within-domain composition matching over full description text. ChemTEB~\citep{shiraee2024chemteb} benchmarks 34 models on chemistry tasks built from filtered QA datasets and finds that contrastive training matters more than domain pretraining; its retrieval evaluation rests on $\sim$33{,}000 corpus passages and 233 queries across two splits adapted from general-domain QA, both substantially smaller than the \fb{} corpus and query set. ChemRAG-Bench~\citep{zhong2025chemrag} evaluates end-to-end RAG for chemistry. DORIS-MAE~\citep{wang2023dorismae} and STaRK~\citep{wu2024stark} highlight that complex structured corpora surface failure modes invisible to standard benchmarks.

\fb{} differs from all prior work in three respects. It is a chemistry retrieval benchmark built on an ontological NER pipeline, enabling graded (0--3) relevance derived from entity-overlap tiers rather than citations or human annotation. Candidate passages are near-misses --- sharing high entity density with the anchor but differing in functional role --- not topic distractors. Finally, commercial API models that lead ChemTEB show no discriminatory signal on \fb{}: task-adaptive fine-tuning on ontologically grounded triplets is required, a gap invisible without structured within-domain negatives.

\section{The \fb{} Dataset}
\label{sec:dataset}

\fb{} is released in three corpus configurations that share identical \texttt{queries.jsonl} and \texttt{qrels/} files, differing only in \texttt{corpus.jsonl}. \textbf{FormBench-Random} provides a $\sim$997K-passage corpus with randomly sampled distractors ($\sim$997K passages). \textbf{FormBench-Structured} provides a $\sim$994K-passage corpus with within-domain distractors drawn from the same chemical macro-domain as each anchor. \textbf{FormBench-Sample} is a compact 63,058-passage subset of scored anchor and negative passages, intended for rapid prototyping. All three variants are distributed in BEIR format under CC-BY-4.0 (Appendix~\ref{app:maintenance}). The full file structure diagram is provided in the dataset card.

\subsection{NER and Ontological Screening}
\label{sec:corpus}

590,247 formulation patents were downloaded from the USPTO via the PatentsView
API~\citep{patentsview}, covering utility patents from 1995--2022. CPC subclass filtering
was applied to retain patents covering pharmaceutics, cosmetics, adhesives, coatings,
agrochemicals, and food formulations. Each patent description was segmented into overlapping chunks (maximum 6,000 characters, 200-character overlap),
producing approximately 3.1 million candidate passages.

Named entity recognition (NER) was applied across 11 entity categories. A Llama-3-8B model with a LoRA adapter~\citep{hu2022lora} was trained via a semi-supervised procedure: Claude Sonnet 4~\citep{anthropic2025sonnet4} generated initial labeled examples from a stratified patent sample, used to fine-tune the adapter across all 11 categories simultaneously. This approach follows
standard practice in materials science NER
literature~\citep{venugopal2022matkg,walker2021matbert,gupta2021matscibert}.
The NER pass processed all 3.1M passages, extracting 234.7M total entity mentions spanning 3,090,473 unique surface forms. Refer to Appendix~\ref{app:ner} for NER details and the annotation schema.

Entity normalization proceeded in three phases: (i)~deterministic normalization ---
lowercasing, punctuation stripping, and plural collapse to merge lexically near-identical
surface forms; (ii)~embedding-based clustering using cosine-similarity thresholding to
group semantically equivalent variants (e.g., ``PEG-400'' and ``polyethylene glycol
400''); and (iii)~LLM adjudication for ambiguous clusters where embedding similarity was
insufficient to resolve cross-category conflicts~\citep{venugopal2022matkg}. The NER
adapter achieved 85.9\% exact-match Jaccard on a 250-passage held-out test set.

\paragraph{Ontological passage similarity.}
After normalization, each passage $P$ maps to a set of canonical entities $\mathcal{E}(P)$. The ontological
similarity between two passages $P_1$ and $P_2$ is their entity-set Jaccard similarity:
\begin{equation}
  J(P_1,\,P_2) \;=\; \frac{|\mathcal{E}(P_1) \cap \mathcal{E}(P_2)|}{|\mathcal{E}(P_1) \cup \mathcal{E}(P_2)|}
  \label{eq:jaccard}
\end{equation}
$J$ ranges from 0 (no shared entities) to 1 (identical entity sets). Passages with
$J \in [0.30,\,0.50]$ are \emph{hard negatives} (HN): compositionally proximate but
functionally distinct. Passages with $J \in [0.05,\,0.15]$ are \emph{soft negatives} (SN):
sharing limited chemical context with the anchor. This tiered structure provides the
graded relevance scores used in \fb{} and the basis for triplet construction in
Section~\ref{sec:queries}.

\subsection{Synthetic Query Generation}
\label{sec:queries}

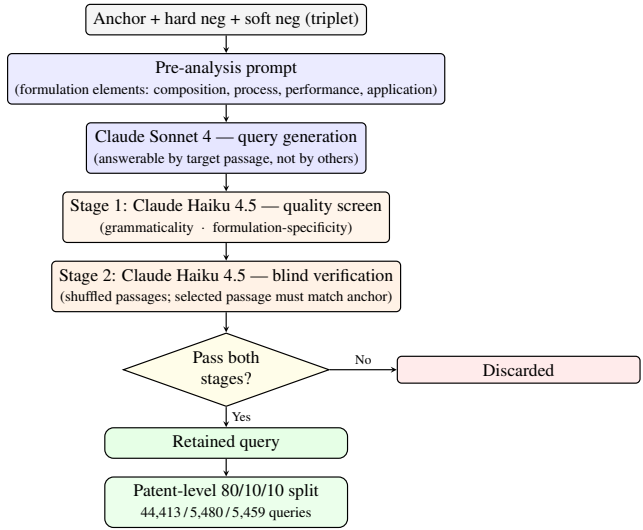
\begin{SCfigure}[0.45][htbp]
  \resizebox{0.60\linewidth}{!}{%
  \begin{tikzpicture}[
    node distance=0.42cm,
    proc/.style={draw, rectangle, rounded corners=3pt, minimum width=5.8cm,
                 minimum height=0.62cm, align=center, font=\large, inner sep=5pt},
    dec/.style={draw, diamond, aspect=2.8, align=center, font=\large,
                minimum width=3.0cm, fill=yellow!10},
    term/.style={draw, rectangle, rounded corners=6pt, minimum width=5.8cm,
                 minimum height=0.62cm, align=center, font=\large, inner sep=5pt},
    >=stealth, thick
  ]
    \node[proc, fill=gray!8]     (pat)     {Anchor + hard neg + soft neg (triplet)};
    \node[proc, fill=blue!8,  below=of pat]   (pre)     {Pre-analysis prompt\\{\normalsize (formulation elements: composition, process, performance, application)}};
    \node[proc, fill=blue!10, below=of pre]   (gen)     {Claude Sonnet 4 --- query generation\\{\normalsize (answerable by target passage, not by others)}};
    \node[proc, fill=orange!8,below=of gen]   (filt)    {Stage 1: Claude Haiku 4.5 --- quality screen\\{\normalsize (grammaticality\enspace$\cdot$\enspace formulation-specificity)}};
    \node[proc, fill=orange!10,below=of filt] (verif)   {Stage 2: Claude Haiku 4.5 --- blind verification\\{\normalsize (shuffled passages; selected passage must match anchor)}};
    \node[dec,                below=0.48cm of verif]    (pass)    {Pass both\\stages?};
    \node[term,fill=green!10, below=0.48cm of pass]     (keep)    {Retained query};
    \node[proc,fill=red!8,    right=1.6cm of pass]      (disc)    {Discarded};
    \node[term,fill=green!8,  below=0.42cm of keep]     (split)   {Patent-level 80/10/10 split\\{\normalsize 44,413\,/\,5,480\,/\,5,459 queries}};

    \draw[->] (pat)   -- (pre);
    \draw[->] (pre)   -- (gen);
    \draw[->] (gen)   -- (filt);
    \draw[->] (filt)  -- (verif);
    \draw[->] (verif) -- (pass);
    \draw[->] (pass)  -- node[right, font=\small] {Yes} (keep);
    \draw[->] (pass)  -- node[above, font=\small] {No}  (disc);
    \draw[->] (keep)  -- (split);
  \end{tikzpicture}}
  \caption{Synthetic query generation pipeline. Triplets pass through Claude Sonnet 4 for
  generation, then two Claude Haiku 4.5 stages for quality screening and blind
  anchor-match verification. Queries failing either stage are discarded.}
  \label{fig:query-gen}
\end{SCfigure}

All \fb{} queries are synthetic, generated by a three-stage LLM pipeline illustrated in
Figure~\ref{fig:query-gen}. Starting from the entity-annotated passage pool, candidate
anchor passages were identified based on entity density. For each anchor, one hard
negative and one soft negative were
retrieved from different patents, forming a triplet. All pairs are cross-patent by
construction.

Each triplet was submitted to Claude Sonnet 4 ~\citep{anthropic2025sonnet4}, which
generated natural-language queries --- concrete, specific questions answerable by a
target passage but not by the other two. Queries were required to address
formulation-relevant content: composition, process conditions, performance properties,
or application context. Questions about pure chemical nomenclature, device hardware, or
patent boilerplate were explicitly excluded. Full generation details are provided in
Appendix~\ref{app:querygen}.

Candidate queries were then subject to two-stage verification by Claude Haiku
4.5~\citep{anthropic2025haiku45}. The first stage screened each query for
grammar and whether it represented a question a formulation scientist would
realistically search for. The second stage presented the query alongside the three passages in
shuffled order and asked the model to identify which passage answers the query. A query
was retained only if the selected passage matched the original anchor; failure at either
stage eliminated it from the qrel pool.

After quality filtering, 55,352 queries were retained and assigned to train/dev/test splits at the \emph{patent} level (80/10/10; 44,413/5,480/5,459 queries; see Appendix~\ref{app:querygen}). All evaluations use the test split (5,459 queries).

We validate both the generated queries and the graded relevance tiers via an LLM-as-judge study on 1,000 stratified tuples (Appendix~\ref{app:qrel-validation}): GPT-4o identifies the anchor as most relevant in 92.6\% of tuples and ranks the hard negative above the soft negative in 64.7\%, both above chance with margin. A domain-expert re-annotation of a 200-tuple sub-sample corroborates the tier structure (A@1 = 78.6\%, HN\,$>$\,SN = 61.1\%, both above chance with margin) and agrees with GPT-4o on the hard-vs-soft ordering in 78.6\% of paired tuples (Appendix~\ref{app:qrel-validation}).

To measure agreement among multiple experts, four practising formulation scientists who were not involved in building the benchmark judged query--passage triples (anchor, hard negative and soft negative in shuffled order). In the first round they completed 72 judgements over 49 triples. They ranked the anchor first in 93.1\% of judgements (95\% CI [84.8, 97.0]), matching GPT-4o on the same triples (91.8\%), and ranked the hard negative above the soft negative in 59.7\% [48.2, 70.3]. Because nearly every judgement agrees on the anchor, chance-corrected statistics such as Cohen's $\kappa$ are uninformative at this prevalence; raw agreement on overlapping judgements is 0.900 and Gwet's AC1 is 0.889. A full study of 100 triples, each judged by more than three experts, is ongoing.

\subsection{Corpus Variants}
\label{sec:variants}

The two full-scale corpus variants differ in how the $\sim$930K distractor slots around
the $\sim$63K anchor and negative passages are filled.

In \textbf{FormBench-Random} (C0), distractor slots are filled by randomly sampling
description chunks from patents not involved in any scored passage pair. The resulting
corpus is topically diverse --- a coatings query competes against passages about
pharmaceutical excipients, food emulsifiers, and rubber compounding --- and measures
coarse-grained domain retrieval.

In \textbf{FormBench-Structured} (C1), distractors are drawn from within the same chemical
macro-domain as the query's anchor. For each anchor passage, all description chunks from
that anchor's patent are added to the corpus. Remaining distractor slots are filled by
sampling patents from the same macro-domain and contributing their complete description
chunks. FormBench-Structured is the primary evaluation configuration;
FormBench-Random provides a controlled comparison that isolates the contribution of
distractor difficulty.

\section{Evaluation Framework}
\label{sec:eval}

\fb{} supports two independent axes of evaluation that can be combined to characterize retrieval quality at different levels of difficulty and granularity.

\paragraph{Axis 1: Corpus difficulty (C0 vs.\ C1).}
The first axis controls the composition of the distractor pool. FormBench-Random (C0) tests whether a model identifies the correct passage among topic-diverse distractors; FormBench-Structured (C1) tests whether a model can discriminate among near-miss passages from the same chemical macro-domain (see Section~\ref{sec:variants} for distractor construction details). The C0/C1 performance gap is itself diagnostic: a large gap indicates that the models have not learned domain-specific discrimination sufficient to distinguish between passages containing similar ontological context.

\paragraph{Axis 2: Evaluation granularity (binary vs.\ graded).}
Consistent with other retrieval benchmarks~\citep{thakur2021beir,muennighoff2023mteb}, we use nDCG@10 as our primary metric:
\begin{equation}
  \mathrm{nDCG@10} = \frac{\mathrm{DCG@10}}{\mathrm{IDCG@10}}, \qquad
  \mathrm{DCG@10} = \sum_{i=1}^{10} \frac{2^{\mathrm{rel}_i} - 1}{\log_2(i+1)}
\end{equation}
where $\mathrm{rel}_i$ is the relevance score of the passage at rank $i$ and $\mathrm{IDCG@10}$ is the ideal (perfectly ranked) DCG. We however use two different types of relevance scoring. \textbf{Binary nDCG@10} treats only the anchor passage (score 3) as relevant and all others as irrelevant (score 0); this matches the evaluation protocol of BEIR and PatenTEB, enabling direct comparison with prior benchmarks. \textbf{Graded nDCG@10} uses the full relevance hierarchy ($\mathrm{rel}_i \in \{0,1,2,3\}$): score 3 for the anchor, score 2 for hard negatives, score 1 for soft negatives, score 0 for all others. Graded evaluation rewards partial credit for retrieving compositionally similar passages near the top of the ranking, revealing finer-grained ordering quality that binary evaluation cannot distinguish. Binary nDCG@10 is the primary reported metric; graded nDCG@10 is reported as secondary. All results are computed on the test split (5,459 queries; \texttt{qrels/test.tsv}).

All metrics are computed with the \texttt{pytrec\_eval} library against the split-specific qrel files. For binary evaluations, the same scoring infrastructure is used with binarized relevance vectors.

\subsection{Evaluation Infrastructure and Latency}
\label{sec:infra}

All evaluations use a two-stage pipeline: corpus embeddings are stored in a FAISS flat inner-product index~\citep{johnson2019faiss} (\texttt{faiss.IndexFlatIP}, $k=100$); at query time each query is encoded and the top-100 results retrieved by exact nearest-neighbor search. We record per-query latency, and embedding throughput. All GPU training and inference was performed on the Frontier supercomputer at Oak Ridge Leadership Computing Facility (OLCF). Full hardware specifications, batch size settings, and per-model throughput figures are in Appendix~\ref{app:experiments}. Commercial API latency reflects round-trip inference time and is not directly comparable to on-device GPU throughput.

\subsection{Models Evaluated}
\label{sec:models-overview}

We evaluate three open-source sentence transformer families---Nomic Embed~\citep{nussbaum2024nomic} (137M), GTE-large~\citep{li2023gte} (434M), and Qwen3-Embedding~\citep{zhang2025qwen3embed} (600M)---under zero-shot and TAPT conditions, and four commercial API baselines---OpenAI \texttt{text-embedding-3-large}/small~\citep{openai2024} and Voyage-4-large/3.5~\citep{voyage2024}---zero-shot only. Full architecture details and training hyperparameters are in Appendix~\ref{app:experiments}.

\section{Experiments}
\label{sec:experiments}

We evaluate the validity and discriminativeness of \fb{} through a structured set of experiments spanning seven zero-shot models and three fine-tuned configurations. The model set is designed to test two questions: (1) do state-of-the-art general-purpose embedders provide any usable retrieval signal on formulation patents without adaptation?; and (2) does task-adaptive contrastive fine-tuning (TAPT) on \fb{}'s own training split provide a learnable signal? These experiments serve as benchmark validation: evidence that \fb{} is discriminative (models differ meaningfully), solvable (at least one configuration achieves non-trivial performance), and that its difficulty is grounded in representational properties that adaptation can partially address. For illustration, five representative query--response examples across three models --- covering API failure, TAPT rescue, a hard case where all models fail, and a within-patent chunk-discrimination failure --- are presented in Appendix~\ref{app:qualitative}.

Table~\ref{tab:main-results} summarises retrieval performance on C1 for all ten model
configurations; detailed analysis per adaptation condition follows in the subsections below.

\subsection{BM25 Lexical Baseline}
\label{sec:bm25}

BM25~\citep{robertson2009bm25} achieves binary \ndcg{} = 0.3751 on C1 and 0.4665 on C0, ranking second on both corpora. As established in BEIR~\citep{thakur2021beir} and corroborated in specialised domains such as legal case retrieval~\citep{rosa2021bm25strong} and patent prior-art retrieval~\citep{althammer2021crossdomain}, lexical baselines remain a competitive reference point that neural retrievers frequently fail to surpass without task-specific adaptation. Since the queries are generated from the passages, lexical overlap between queries and anchor passages is expected, and the strong BM25 performance is not surprising. BM25 therefore provides a useful baseline against which to assess how learned embeddings perform relative to simpler lexical matching.

\subsection{Base Model Results}
\label{sec:zero-shot}

Base model results reveal a sharp divide. The four commercial APIs score near-zero on C1
(binary \ndcg{}: Voyage-3.5: 0.0025, Voyage-4-large: 0.0064, OpenAI-3-small: 0.0079,
OpenAI-3-large: 0.0090); the 0.0011 gap between OpenAI large and small despite a
$2\times$ dimensionality difference confirms that scaling general-purpose capacity provides
no benefit here. The three open-source base models achieve substantially higher binary
\ndcg{}: Nomic 0.1475, GTE-large 0.2056, Qwen3-Embed 0.2317.
None of the base models, however, approach the BM25 lexical baseline (0.3751 on C1):
even the strongest open-source encoder (Qwen3-Embed) trails BM25 by 0.144.

\subsection{TAPT}
\label{sec:tapt}

\subsubsection{Protocol}

\textbf{Task-Adaptive Pre-Training (TAPT)} fine-tunes the model contrastively on the \fb{} training split using Multiple Negatives Ranking Loss~\citep{henderson2017mnrl,oord2018infonce} (MNRL). Each training example is a triplet $(Q,\,A,\,\text{HN})$ where $Q$ is the generated query, $A$ is the anchor passage (score 3), and $\text{HN}$ is a hard negative from the Jaccard-tier pairs (score 2). MNRL treats $A$ as the positive for $Q$, uses $\text{HN}$ as an explicit negative, and additionally treats the other in-batch positives as negatives. MNRL directly trains the geometry of the embedding space, moving matched passages closer to queries and pushing hard negatives apart.

\subsubsection{Results}
\label{sec:tapt-results}

TAPT-MNRL on the \fb{} training split produces substantial retrieval improvements across all three architectures. Contrastive fine-tuning with task-specific $(Q,\,A,\,\text{HN})$ triplets directly reshapes the embedding geometry in a way that MLM pretraining cannot~\citep{karpukhin2020dpr,gao2021simcse}. Qwen3-Embedding + MNRL achieves the best result at binary \ndcg{} = 0.4085 on C1 (graded \ndcg{} = 0.2374). Nomic + MNRL reaches binary \ndcg{} = 0.3668 and GTE + MNRL reaches 0.3561. Qwen3-Embed is the only TAPT configuration to exceed the BM25 lexical baseline (0.4085 vs.\ 0.3751 on C1; 0.4835 vs.\ 0.4665 on C0); Nomic and GTE approach but do not surpass it, suggesting that model capacity determines whether contrastive fine-tuning can overcome the lexical ceiling. The magnitude of improvement is consistent with the results of contrastive task-adaptive fine-tuning in other specialized domains~\citep{tang2024finmteb,shiraee2024chemteb}. Hard-negative quality is a critical determinant of contrastive fine-tuning performance~\citep{moreira2024nvretriever}; \fb{}'s taxonomy-derived triplets pair each query with semantically close negatives from the same formulation cluster. Full per-checkpoint results are in Appendix~\ref{app:training}. TAPT also transfers to a patent-domain encoder: PaECTER + MNRL rises from 0.1334 to 0.2338 on C1 (1.75$\times$), and to 0.2486 when passages are chunk-pooled rather than truncated, while its untuned base remains well below the general open-source encoders.

\begin{table}[htbp]
  \caption{Retrieval results on \fb{}-Structured (C1) and \fb{}-Random (C0) test sets ($n$=5,459).
  \ndcg{} is reported under binary qrels (anchor=1, others=0; BEIR-comparable) and graded qrels (tier scores 0--3).
  Latency: FAISS ms/query for neural models; BM25 retrieval ms/query for lexical baseline (encoding excluded). Best per column in \textbf{bold}.
  Patent encoders truncate passages at their context window, except rows marked *, which encode each passage in $\le$500-token windows and mean-pool (no truncation; C1 only).
  \dag~From a separate run whose truncated references are 0.1334 (PaECTER) and 0.1064 (PatentSBERTa).}
  \label{tab:main-results}
  \centering
  \small
  \begin{tabular}{llrrrrr}
    \toprule
    & & \multicolumn{2}{c}{\textbf{nDCG@10}} & & & \\
    \cmidrule(lr){3-4}
    \textbf{Model} & \textbf{Config} &
    \textbf{Binary} & \textbf{Graded} &
    \textbf{MRR} & \textbf{R@100} & \textbf{ms/q} \\
    \midrule
    \multicolumn{7}{l}{\textit{\fb{}-Structured (C1)}} \\
    \midrule
    BM25            & Lexical & 0.3751 & 0.2216 & 0.3252 & 0.8368 & 11.7 \\
    OpenAI-3-large  & Base & 0.0090 & 0.0058 & 0.0077 & 0.0445 & 553.8 \\
    OpenAI-3-small  & Base & 0.0079 & 0.0053 & 0.0065 & 0.0366 & 284.5 \\
    Voyage-4-large  & Base & 0.0064 & 0.0043 & 0.0053 & 0.0343 & 192.5 \\
    Voyage-3.5      & Base & 0.0025 & 0.0016 & 0.0020 & 0.0212 & 190.6 \\
    Nomic           & Base & 0.1475 & 0.0881 & 0.1308 & 0.4076 & 146.8 \\
    GTE-large       & Base & 0.2056 & 0.1245 & 0.1798 & 0.5638 & 191.7 \\
    Qwen3-Embed     & Base & 0.2317 & 0.1384 & 0.2031 & 0.5721 & 191.3 \\
    \midrule
    \multicolumn{7}{l}{\textit{Patent-domain encoders}} \\
    patembed-long4096 & Base & 0.1783 & 0.1078 & 0.1546 & 0.4999 & 157.9 \\
    PaECTER           & Base & 0.1329 & 0.0808 & 0.1154 & 0.3810 & 202.3 \\
    PaECTER           & Base\textsuperscript{*\dag} & 0.1239 & -- & -- & -- & -- \\
    PatentSBERTa      & Base & 0.1065 & 0.0639 & 0.0932 & 0.2925 & 156.1 \\
    PatentSBERTa      & Base\textsuperscript{*\dag} & 0.0963 & -- & -- & -- & -- \\
    patembed-large    & Base & 0.0665 & 0.0418 & 0.0581 & 0.2438 & 201.6 \\
    PaECTER           & TAPT\textsuperscript{\dag} & 0.2338 & -- & -- & -- & -- \\
    PaECTER           & TAPT\textsuperscript{*\dag} & 0.2486 & -- & -- & -- & -- \\
    \midrule
    Nomic           & TAPT & 0.3668 & 0.2145 & 0.3228 & 0.7903 & 147.0 \\
    GTE-large       & TAPT & 0.3561 & 0.2087 & 0.3125 & 0.7586 & 195.0 \\
    \textbf{Qwen3}  & \textbf{TAPT} & \textbf{0.4085} & \textbf{0.2374} & \textbf{0.3613} & \textbf{0.8181} & 190.6 \\
    \midrule
    \multicolumn{7}{l}{\textit{\fb{}-Random (C0)}} \\
    \midrule
    BM25            & Lexical & 0.4665 & 0.2776 & 0.4173 & 0.8652 & 13.3 \\
    OpenAI-3-large  & Base & 0.2132 & 0.1279 & 0.1880 & 0.5677 & 554.4 \\
    OpenAI-3-small  & Base & 0.1985 & 0.1192 & 0.1738 & 0.5283 & 285.0 \\
    Voyage-4-large  & Base & 0.1832 & 0.1094 & 0.1578 & 0.5924 & 192.3 \\
    Voyage-3.5      & Base & 0.0529 & 0.0309 & 0.0455 & 0.3074 & 191.4 \\
    Nomic           & Base & 0.1658 & 0.1004 & 0.1482 & 0.4263 & 146.9 \\
    GTE-large       & Base & 0.2420 & 0.1476 & 0.2164 & 0.5917 & 191.6 \\
    Qwen3-Embed     & Base & 0.2652 & 0.1596 & 0.2387 & 0.5812 & 192.3 \\
    \midrule
    \multicolumn{7}{l}{\textit{Patent-domain encoders}} \\
    patembed-long4096 & Base & 0.2199 & 0.1345 & -- & -- & -- \\
    PaECTER           & Base & 0.1640 & 0.1006 & -- & -- & -- \\
    PatentSBERTa      & Base & 0.1209 & 0.0727 & -- & -- & -- \\
    patembed-large    & Base & 0.0805 & 0.0516 & -- & -- & -- \\
    \midrule
    Nomic           & TAPT & 0.4358 & 0.2583 & 0.3915 & 0.8311 & 146.6 \\
    GTE-large       & TAPT & 0.4308 & 0.2553 & 0.3883 & 0.8011 & 191.0 \\
    \textbf{Qwen3}  & \textbf{TAPT} & \textbf{0.4835} & \textbf{0.2835} & \textbf{0.4404} & \textbf{0.8525} & 191.0 \\
    \bottomrule
  \end{tabular}
\end{table}

\subsection{C0 vs.\ C1 Corpus Structure}
\label{sec:c0c1-exp}

\begin{SCfigure}[0.45][!t]
  \includegraphics[width=0.65\linewidth]{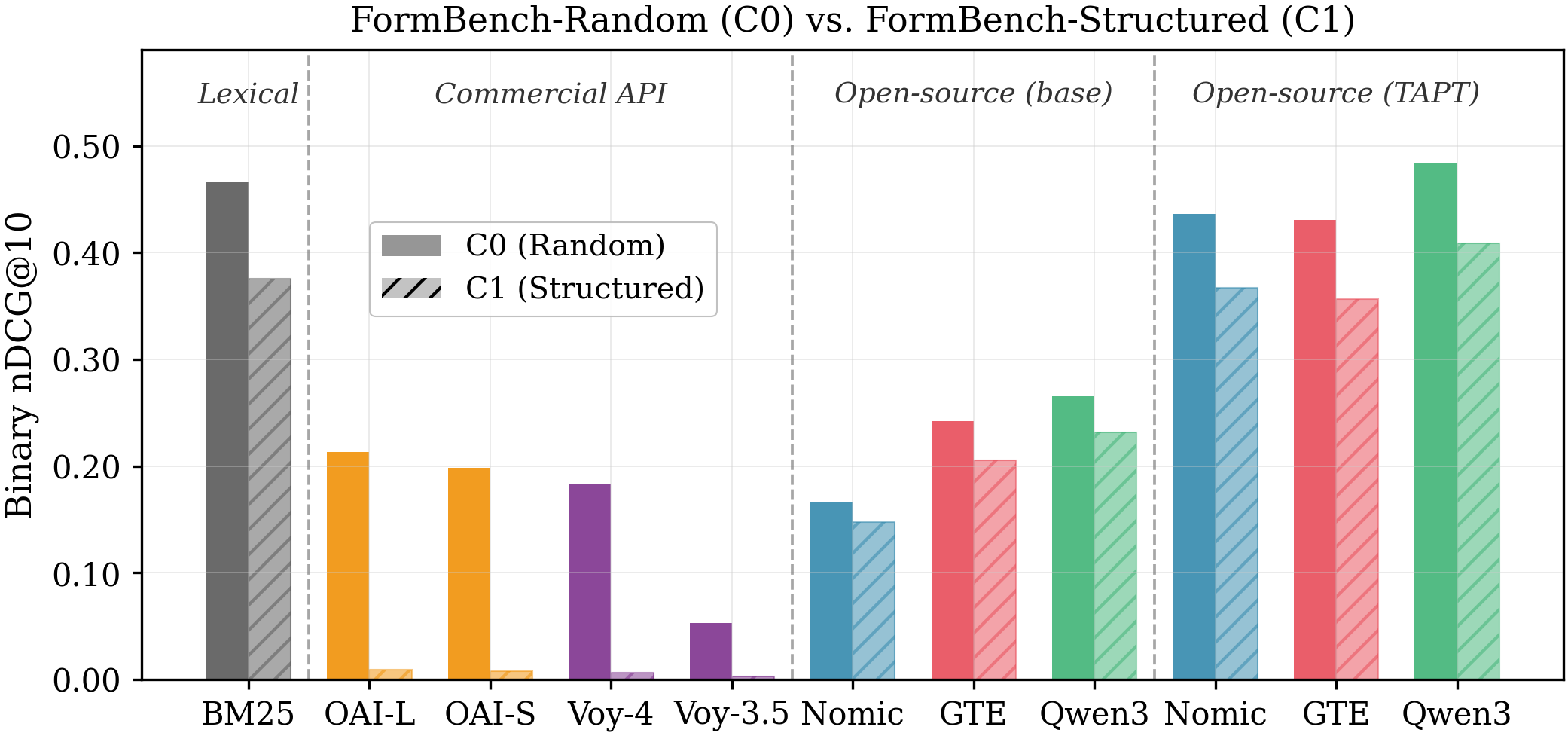}
  \caption{Binary \ndcg{} on C0 (random distractors) and C1 (structured near-miss distractors) for all models, including the BM25 lexical baseline. The API/open-source gap is far larger on C1; TAPT models show a near-uniform C0/C1 $\Delta$ ($\approx$0.07), reflecting composition-level discrimination; BM25 shows a larger $\Delta$ (0.091) as structured within-domain distractors suppress lexical signal.}
  \label{fig:c0c1}
\end{SCfigure}

The C0/C1 gap diagnoses what each method depends on. API models depend on coarse domain matching: OpenAI-3-large scores 0.213 on C0 but collapses to 0.009 on C1 ($\Delta = +0.204$) when random distractors are replaced by structured near-miss passages. BM25 depends on lexical overlap: within-domain distractors share patent vocabulary with queries, suppressing its core signal ($\Delta = 0.091$). TAPT models depend on neither: the gap is small and uniform ($\Delta \approx 0.070$--$0.075$ across all three architectures), indicating that contrastive fine-tuning builds composition-level discrimination rather than domain-level pattern matching (Figure~\ref{fig:c0c1}).

Four patent-domain encoders, pretrained on patent text, behave differently from the APIs (Table~\ref{tab:main-results}). They start lower than the general open-source models (binary \ndcg{} on C1: 0.0665--0.1783) but retain 81--88\% of their C0 score on C1, against 4\% for OpenAI-3-large. Removing truncation does not help: chunk-and-pool encoding lowers PaECTER from 0.1334 to 0.1239 and PatentSBERTa from 0.1064 to 0.0963, and the 4,096-token patembed-long4096 is the strongest of the four. Patent pretraining therefore buys robustness to structured distractors but not fine-grained formulation discrimination.

\subsection{Corpus-Size Sensitivity}
\label{sec:scaling-exp}

\begin{SCfigure}[0.45][!t]
  \includegraphics[width=0.65\linewidth]{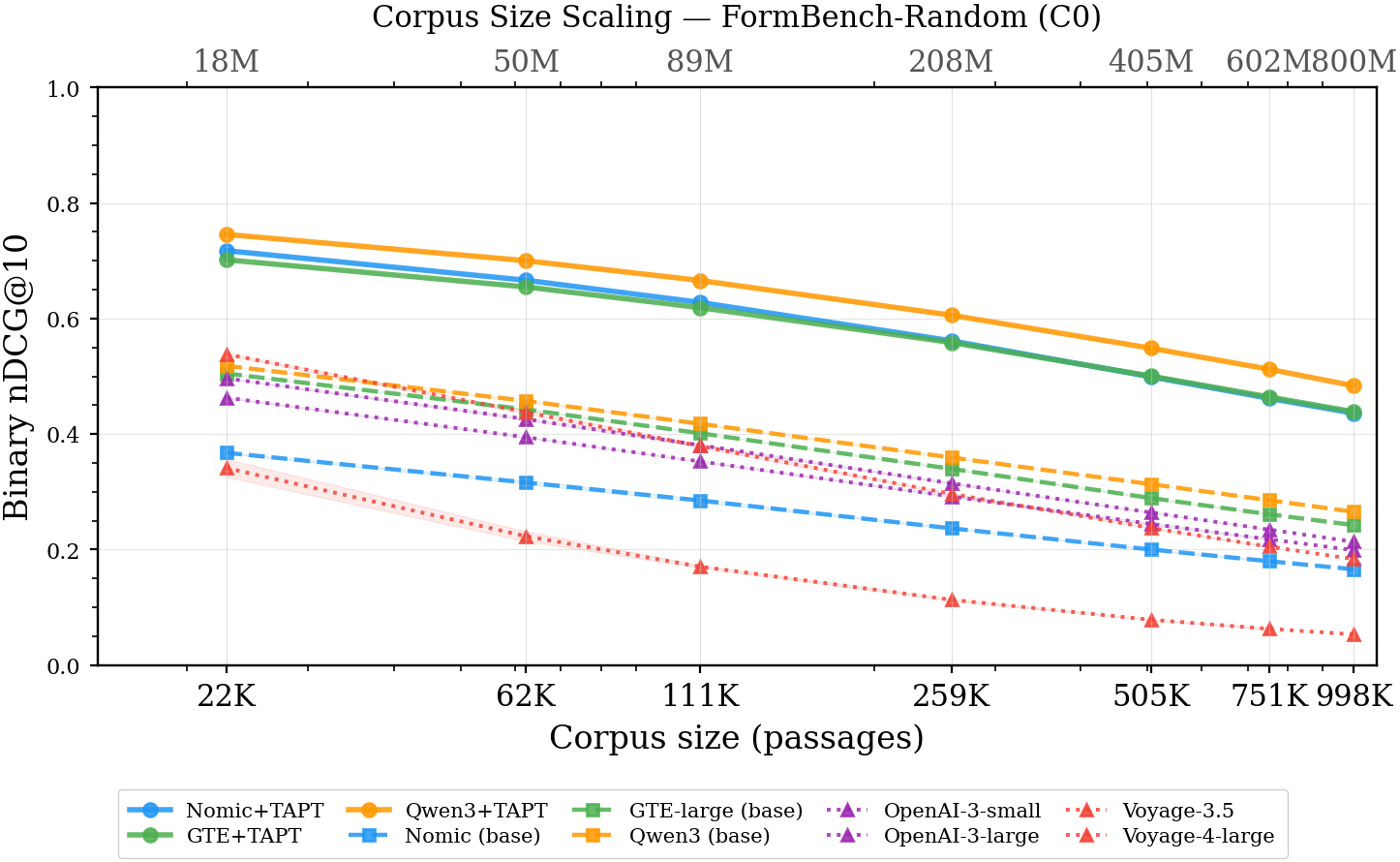}
  \caption{Binary \ndcg{} vs.\ C0 distractor corpus fraction (mean $\pm$1 s.d., 3 seeds, 10 models). API models collapse fastest (Voyage-3.5: $-82\%$; OpenAI-3-large: $-53\%$ from $f$=0.01 to 0.75). TAPT models degrade slowest (Qwen3+MNRL: $-31\%$).}
  \label{fig:corpus-size}
\end{SCfigure}

To test whether API performance degrades faster than open-source models as corpus size grows, we evaluate all 10 models on subsampled C0 corpora spanning six distractor fractions ($\{0.01, 0.05, 0.10, 0.25, 0.50, 0.75\}$, three seeds each, $\sim$23K to $\sim$749K distractor passages). All models degrade with increasing corpus size --- this is a general effect of a harder retrieval haystack as shown in Figure~\ref{fig:corpus-size}. The key difference is the rate of degradation. At small corpus sizes ($0.01$), API models score comparably to or above open-source zero-shot baselines: OpenAI-3-large reaches binary nDCG@10\,=\,0.497, Voyage-4-large 0.538. As the corpus grows, API scores collapse sharply: Voyage-3.5 falls from 0.341 at fraction 0.01 to 0.063 at fraction 0.75 (an 82\% drop); OpenAI-3-large falls from 0.497 to 0.235 (53\%). Open-source base models show moderate degradation: Qwen3-base from 0.518 to 0.285 (45\%). TAPT-fine-tuned models degrade most slowly: Qwen3+MNRL from 0.746 to 0.512 (31\%), Nomic+MNRL from 0.718 to 0.462 (36\%). The interpretation is consistent with the geometric framing: at small corpus sizes, coarse domain matching is sufficient for reasonable retrieval; as the corpus grows, the discriminative signal required to separate near-miss passages within the same domain becomes critical --- a capability that only contrastive fine-tuning provides.

Full results across all model configurations, metrics, and corpus variants are in Appendix~\ref{app:experiments}.

\section{Analysis}
\label{sec:analysis}

The experimental results in Section~\ref{sec:experiments} establish a clear hierarchy: commercial API embeddings fail on \fb{} while TAPT-fine-tuned open-source models are competitive or outperform the baseline. Two deeper questions motivate the analysis that follows. First, \emph{why} do massively pre-trained API embeddings---state-of-the-art on general benchmarks---score near zero on this task? The answer cannot simply be domain unfamiliarity, since these same models perform well on coarse patent classification (Section~\ref{sec:downstream}). Second, what does TAPT training actually do to the geometry of the embedding space that enables successful retrieval? We examine both questions through a diagnostic geometric analysis and downstream task probes.

\subsection{Embedding Geometry: Why Commercial APIs Fail}
\label{sec:geometry}

\begin{figure}[htbp]
  \centering
  \begin{minipage}[c]{0.57\linewidth}
    \centering
    \includegraphics[width=\linewidth]{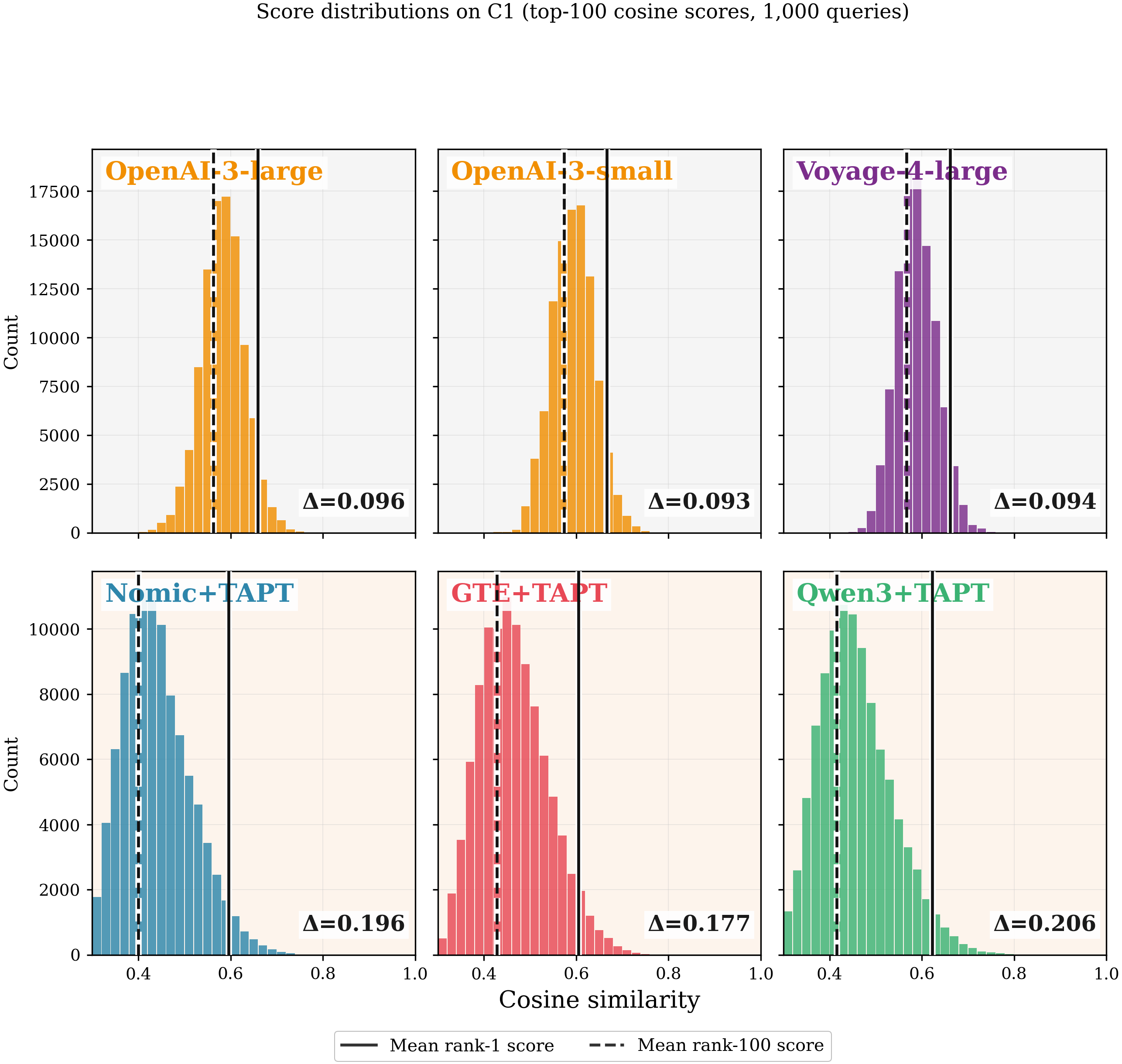}
    \caption{Top-100 cosine similarity distributions on the C1 corpus (1,000 sampled queries).
    $\Delta$ = mean rank-1 minus mean rank-100 score. API models collapse to a narrow
    band ($\Delta \approx 0.09$); TAPT models achieve the widest separation
    ($\Delta = 0.177$--$0.206$ across the three architectures).}
    \label{fig:score-dist}
  \end{minipage}
  \hfill
  \begin{minipage}[c]{0.39\linewidth}
    \centering
    \captionof{table}{Percentage of test queries ($n$=5,459) for which the anchor
    passage is ranked first (Anchor@1) among 100 FAISS candidates from the C1 corpus. Best in \textbf{bold}.}
    \label{tab:recall-k}
    \small
    \begin{tabular}{lr}
      \toprule
      \textbf{Model (Type)} & \textbf{@1 (\%)} \\
      \midrule
      OpenAI-3-large (API)       &  0.3 \\
      OpenAI-3-small (API)       &  0.3 \\
      Voyage-4-large (API)       &  0.2 \\
      Voyage-3.5 (API)           &  0.0 \\
      \midrule
      Nomic (Base)               &  8.3 \\
      GTE-large (Base)           & 10.9 \\
      Qwen3 (Base)               & 12.9 \\
      \midrule
      GTE-large (TAPT)           & 20.4 \\
      Nomic (TAPT)               & 21.0 \\
      \textbf{Qwen3 (TAPT)}      & \textbf{24.5} \\
      \bottomrule
    \end{tabular}
  \end{minipage}
\end{figure}

To understand the geometric failure of commercial API models, we run a diagnostic experiment: for each model, we extract the per-query cosine similarity distribution over the top-100 retrieved passages from the C1 corpus for all 5,459 test queries. If the model has learned to discriminate within the formulation patent domain, we expect the ``discrimination gap'' $\Delta$ between the top-ranked and rank-100 score to be large. If the embedding space collapses all within-domain passages to a narrow similarity band, $\Delta$ will approach zero. Figure~\ref{fig:score-dist} shows these distributions for representative models.

The result is unambiguous. Commercial API models assign a moderate similarity score to essentially every passage: $\Delta = 0.096$ for OpenAI text-embedding-3-large---the embedding space has correctly identified formulation chemistry as a coherent domain, but cannot discriminate \emph{within} it. This is the within-domain anisotropy problem: general embedders compress specialised text into a narrow cone where pairwise similarities are uniformly high, leaving no usable ranking signal~\citep{shiraee2024chemteb,wang2023dorismae}. TAPT-fine-tuned models show $\Delta = 0.177$--$0.206$ across the three architectures---a wider discrimination band---because contrastive training on hard negatives directly optimises pairwise separation~\citep{gao2021simcse}, pushing non-matching passages downward in score space.

This is further explained by the anchor rank distributions in Table~\ref{tab:recall-k}. OpenAI text-embedding-3-large places the anchor at rank~1 for only 0.3\% of test queries. Qwen3-Embed TAPT-MNRL, by contrast, achieves 24.5\%. The zero-shot open-source models occupy an intermediate position, confirming that contrastive pretraining provides partial but insufficient within-domain discrimination. TAPT directly reshapes the embedding space so that the anchor consistently outranks its near-miss distractors.

\subsection{Downstream Task Evaluation: What TAPT Does and Does Not Learn}
\label{sec:downstream}

We test whether TAPT gains transfer to CPC classification and clustering (following the MTEB evaluation paradigm~\citep{muennighoff2023mteb}), using passage-level CPC labels fully disjoint from TAPT supervision. All USPTO patents carry CPC codes; we use the coarse 7-class section labels (A = Human Necessities, C = Chemistry, etc.) and run logistic regression probes on frozen embeddings (40,005 train / 4,997 test) and bootstrapped V-measure clustering.

The result is a clean inversion (Table~\ref{tab:cls-clu}; UMAP
projections in Appendix~\ref{app:clustering}). API models lead on classification:
OpenAI-3-large achieves 80.4\% accuracy; the best TAPT model
(GTE-large) reaches 78.0\%---a 2.4-point gap. TAPT adds no
improvement over zero-shot base models on either task. While this confirms that the
capabilities are orthogonal, the exact mechanism is unclear. The pattern is consistent
with what other domain-specific benchmarks have reported across
chemistry~\citep{shiraee2024chemteb}, patents~\citep{ayaou2025patenteb}, and legal
text~\citep{butler2025mleb}. Representative UMAP projections illustrating the geometry of
each model family are provided in Appendix~\ref{app:clustering}.

\begin{table}[htbp]
  \caption{CPC classification (frozen logistic regression, 7 sections) and clustering (V-measure, 10 bootstrap samples) for all 10 models. API models lead on both tasks despite near-zero C1 retrieval; TAPT provides no improvement, confirming retrieval and taxonomic encoding are orthogonal. 80-class results in Appendix~\ref{app:clustering}. Best per column in \textbf{bold}.}
  \label{tab:cls-clu}
  \centering
  \small
  \begin{tabular}{llrrr}
    \toprule
    \textbf{Model} & \textbf{Config} & \textbf{CLS Acc} & \textbf{CLS F1-mac} & \textbf{CLU V} \\
    \midrule
    \textbf{OpenAI-3-large} & Base & \textbf{0.804} & 0.522 & 0.191 \\
    OpenAI-3-small          & Base & 0.785          & 0.475 & 0.174 \\
    Voyage-4-large          & Base & 0.802          & \textbf{0.530} & 0.188 \\
    Voyage-3.5              & Base & 0.788          & 0.521 & \textbf{0.208} \\
    \midrule
    Nomic                   & Base & 0.750          & 0.423 & 0.149 \\
    GTE-large               & Base & 0.770          & 0.458 & 0.165 \\
    Qwen3-Embed             & Base & 0.778          & 0.477 & 0.166 \\
    \midrule
    Nomic                   & TAPT (MNRL) & 0.758   & 0.463 & 0.162 \\
    GTE-large               & TAPT (MNRL) & 0.780   & 0.474 & 0.175 \\
    Qwen3                   & TAPT (MNRL) & 0.772   & 0.462 & 0.171 \\
    \bottomrule
  \end{tabular}
\end{table}

\subsection{Analysis of Lexical Similarity}
\label{sec:lexical}

Because each query is generated from its anchor passage, part of a model's score may come from shared vocabulary rather than understanding. We test this directly. GPT-4o, a different model family from the Claude query generator, rewrites 1,500 C1 test queries over up to three rounds, giving four paired levels (Table~\ref{tab:lexical}). Rewriting lowers the IDF-weighted share of each query's vocabulary found in its anchor from 0.687 to 0.264 ($-62\%$), and in the hard negative from 0.282 to 0.127 ($-55\%$). BM25 falls 77\% (0.3765 $\rightarrow$ 0.0853), dropping from third to sixth place, while all six neural models fall within a narrow 43--52\% band; TAPT-Qwen3's lead over BM25 therefore grows from 1.10$\times$ to 2.53$\times$. The uniform neural drop reflects a common cost of the rewritten queries; the further loss in BM25 is attributable to its reliance on exact word overlap. OpenAI-3-large stays near zero at every level (0.0089 $\rightarrow$ 0.0060), so its failure on C1 is not an artifact of lexical overlap. Even on unmodified queries, BM25 depends strongly on natural overlap: bucketed by query--anchor overlap quartile, its score rises from 0.1920 to 0.5076. An independent check that the rewritten queries preserve their meaning has not yet been run, so part of the drop may reflect queries that lost information.

\begin{table}[htbp]
  \caption{Binary \ndcg{} on C1 under progressive paraphrasing (1,500 paired test queries).
  L100: original; Lbare: one GPT-4o rewrite; L75/L50: rewritten until IDF-weighted
  query--anchor overlap falls to $\le$75\%/$\le$50\% of its original value. Our TAPT models in \textbf{bold}.}
  \label{tab:lexical}
  \centering\small
  \begin{tabular}{llrrrrr}
    \toprule
    \textbf{Model} & \textbf{Config} & \textbf{L100} & \textbf{Lbare} & \textbf{L75} & \textbf{L50} & $\Delta$ \\
    \midrule
    BM25           & Lexical & 0.3765 & 0.2937 & 0.1348 & 0.0853 & $-77\%$ \\
    OpenAI-3-large & Base    & 0.0089 & 0.0091 & 0.0065 & 0.0060 & $-33\%$ \\
    Nomic          & Base    & 0.1533 & 0.1424 & 0.0963 & 0.0734 & $-52\%$ \\
    GTE-large      & Base    & 0.2109 & 0.1918 & 0.1436 & 0.1166 & $-45\%$ \\
    Qwen3-Embed    & Base    & 0.2381 & 0.2203 & 0.1550 & 0.1229 & $-48\%$ \\
    \textbf{Nomic}     & \textbf{TAPT} & 0.3803 & 0.3498 & 0.2414 & 0.1933 & $-49\%$ \\
    \textbf{GTE-large} & \textbf{TAPT} & 0.3691 & 0.3380 & 0.2570 & 0.2107 & $-43\%$ \\
    \textbf{Qwen3}     & \textbf{TAPT} & 0.4149 & 0.3721 & 0.2665 & 0.2161 & $-48\%$ \\
    \bottomrule
  \end{tabular}
\end{table}

\subsection{External Validation}
\label{sec:external}

We evaluate our models on two external chemistry retrieval benchmarks (Table~\ref{tab:external}). The retrieval component of ChemTEB~\citep{shiraee2024chemteb} consists of ChemNQ and ChemHotpotQA, with 45 test queries in total; this is too small to draw conclusions from, and we report these results for completeness. ChemRxivRetrieval (\texttt{BASF-AI/ChemRxivRetrieval}), released by BASF with the ChEmbed models~\citep{shiraee2025chembed}, pairs 5,000 LLM-generated queries with 69,457 passages from ChemRxiv papers, with one relevant passage per query.

On ChemRxivRetrieval our pipeline reproduces the published scores to within 0.001 for three of four reference models (nomic-v1 0.8208 vs.\ 0.821; nomic-v1.5 0.7832 vs.\ 0.783; Qwen3-Embed 0.8198 vs.\ 0.819; OpenAI-3-large 0.7802 vs.\ 0.775), independently validating the scorer behind every \fb{} number. TAPT on \fb{} slightly improves both models tested (Nomic 0.7832 $\rightarrow$ 0.7949; Qwen3 0.8198 $\rightarrow$ 0.8289; not significance-tested), and Qwen3-TAPT is the strongest model we ran, though below the ChEmbed models trained on ChemRxiv data (0.911 and 0.902, published). Conversely, the chemistry-specialised ChEmbed models are the strongest base models we evaluated on \fb{} (ChEmbed-prog 0.2456 on C1, 0.2881 on C0), yet remain below BM25 and 1.66$\times$ below Qwen3-TAPT: broad chemistry training does not solve formulation retrieval. Together, these results validate our evaluation externally and show that \fb{} measures a capability that existing chemistry benchmarks and models do not cover.

\begin{table}[htbp]
  \caption{External chemistry retrieval benchmarks (\ndcg{}). ChemTEB retrieval comprises
  ChemNQ and ChemHotpotQA (45 test queries in total) and is too small to support conclusions.
  ChEmbed values are as published. Our TAPT models in \textbf{bold}.}
  \label{tab:external}
  \centering\small
  \begin{tabular}{llrrrr}
    \toprule
    & & \multicolumn{3}{c}{\textbf{ChemTEB}} & \textbf{ChemRxiv-} \\
    \cmidrule(lr){3-5}
    \textbf{Model} & \textbf{Config} & \textbf{ChemNQ} & \textbf{HotpotQA} & \textbf{Mean} & \textbf{Retrieval} \\
    \textit{Queries / docs} & & \textit{27 / 22,933} & \textit{18 / 10,069} & & \textit{5,000 / 69,457} \\
    \midrule
    OpenAI-3-large  & Base & 0.6254 & 0.8044 & 0.7149 & 0.7802 \\
    Nomic           & Base & 0.6211 & 0.7815 & 0.7013 & 0.7832 \\
    GTE-large       & Base & 0.6441 & 0.7142 & 0.6791 & -- \\
    Qwen3-Embed     & Base & 0.6380 & 0.7800 & 0.7090 & 0.8198 \\
    \textbf{Nomic}     & \textbf{TAPT} & 0.5173 & 0.7402 & 0.6288 & 0.7949 \\
    \textbf{GTE-large} & \textbf{TAPT} & 0.6094 & 0.7295 & 0.6695 & -- \\
    \textbf{Qwen3}     & \textbf{TAPT} & 0.5536 & 0.7876 & 0.6706 & 0.8289 \\
    ChEmbed-prog    & Published & -- & -- & -- & 0.911 \\
    ChEmbed-vanilla & Published & -- & -- & -- & 0.902 \\
    \bottomrule
  \end{tabular}
\end{table}

\section{Conclusion}
\label{sec:conclusion}

We introduce \fb{}, a large-scale retrieval benchmark of US formulation patents that exposes a failure mode invisible to existing benchmarks: the inability of general-purpose embeddings to discriminate between chemically similar passages within the same domain. Commercial API models fail structurally---OpenAI \texttt{text-embedding-3-large} achieves binary \ndcg{} = 0.009 on C1, scoring below a BM25 lexical baseline (0.375 on C1, 0.467 on C0). TAPT on structured hard negatives recovers the gap: Qwen3-Embed + MNRL reaches 0.4085 on C1 and 0.4835 on C0, the only configuration to exceed the BM25 floor. The same API models that fail at retrieval lead on CPC classification probes. This retrieval--classification dissociation provides a reusable diagnostic: when classification succeeds where retrieval fails, TAPT on structured hard negatives is the appropriate fix. \fb{} is released in BEIR format under CC-BY-4.0.

\paragraph{Limitations.}
Relevance judgments are a computational proxy (Jaccard over entity sets). The graded qrel tiers, validated via LLM-as-judge at 64.7\% hard-over-soft agreement (Appendix~\ref{app:qrel-validation}), provide signal above chance but are not a precise proxy for human relevance judgements at fine granularity; binary \ndcg{} (score-3 vs.\ all others) is more robust to tier-boundary uncertainty and is our primary reported metric. Coverage is limited to USPTO utility patents (1995--2022) in English only.

\paragraph{Future work.} Replacing the LLM-synthesized queries with human-authored queries from formulation chemists would directly address the principal validity limitation. Extending coverage beyond USPTO English-only patents to multilingual and non-US sources would broaden applicability.

\paragraph{Maintenance.} Release, hosting, versioning, and evaluation tooling are detailed in Appendix~\ref{app:maintenance}.

\begin{ack}

We thank the Oak Ridge Leadership Computing Facility (OLCF) for compute support on Frontier.
This research used resources of OLCF, a U.S.\ Department of Energy Office of Science User
Facility supported under Contract DE-AC05-00OR22725.
We thank Albert Invent for API access and financial support.

\end{ack}

\section*{References}
\bibliographystyle{unsrtnat}
\bibliography{references}

\appendix

\section{Dataset Construction Details}
\label{app:construction}

\subsection{Patent Corpus and NER Pipeline}
\label{app:ner}

\paragraph{Patent corpus acquisition.}
590,247 US utility patents were downloaded from the USPTO via the PatentsView bulk data API, covering the grant years 1995--2022. CPC subclass filtering retained patents assigned to subclasses associated with pharmaceutical dosage forms (A61K, A61P), adhesives (C09J), coatings and paints (C09D), agricultural formulations (A01N), food compositions (A23L), and cosmetics (A61Q). Each patent description was segmented at paragraph boundaries into overlapping chunks of at most 6,000 characters with a 200-character overlap, producing 3,097,441 candidate passages. Passages shorter than 200 characters or containing no alphabetic content were discarded, leaving approximately 3.1 million passages for downstream processing.

\paragraph{NER model.}
Named entity recognition was performed using a Llama-3-8B model with a LoRA adapter fine-tuned simultaneously across the 11 entity categories listed in Table~\ref{tab:ner-counts}. Training data was constructed via a semi-supervised procedure: Claude Sonnet 4~\citep{anthropic2025sonnet4} generated 25,000 annotated passages from a stratified patent sample (proportional to CPC macro-domain size); these were reviewed for consistency, deduplicated, and used to fine-tune the LoRA adapter for 3 epochs at learning rate 5$\times$10$^{-4}$. The adapter achieved 85.9\% exact-match Jaccard on a 250-passage held-out test set. The full 3.1M-passage pool was then processed in parallel on OLCF Frontier, extracting 234.7M entity mentions.

\paragraph{Entity normalisation.}
Raw NER output was normalised in three phases. (i)~\emph{Deterministic normalisation}: lowercasing, punctuation stripping (excluding chemical notation characters), plural collapse, and unit standardisation (e.g., ``\% w/w'' normalised to ``\%w/w'') to merge lexically near-identical surface forms. (ii)~\emph{Embedding-based clustering}: OpenAI \texttt{text-embedding-3-large} embeddings were computed for all unique surface forms; forms within cosine similarity $>0.95$ were merged to their centroid representative, resolving variants such as ``PEG-400'' / ``polyethylene glycol 400'' / ``PEG400''. (iii)~\emph{LLM adjudication}: 12,847 ambiguous clusters (where embedding similarity was insufficient to resolve cross-category conflicts or chemical synonymy) were submitted to Claude Haiku 4.5~\citep{anthropic2025haiku45} for binary merge/split decisions. After normalisation, the 3,090,473 unique raw surface forms collapsed to 2,854,881 canonical forms (Table~\ref{tab:ner-counts}).

\begin{table}[h]
  \caption{Unique entity surface forms before and after normalization across the full
  3.1M-passage pool (11 NER categories; 234.7M total mentions).}
  \label{tab:ner-counts}
  \centering
  \small
  \begin{tabular}{lrr}
    \toprule
    \textbf{Entity Type} & \textbf{Unique Raw} & \textbf{Unique Canonical} \\
    \midrule
    Materials      & 939,484   & 846,690 \\
    Values         & 1,007,337 & 931,758 \\
    Processes      & 206,737   & 195,570 \\
    Components     & 345,388   & 324,762 \\
    Applications   & 337,033   & 311,200 \\
    Properties     & 141,303   & 133,719 \\
    Roles          & 71,182    & 64,239  \\
    Conditions     & 145,646   & 139,347 \\
    Equipment      & 150,792   & 142,086 \\
    Devices        & 151,987   & 140,546 \\
    Sectors        & 13,079    & 12,897  \\
    \midrule
    \textbf{Total} & \textbf{3,090,473} & \textbf{2,854,881} \\
    \bottomrule
  \end{tabular}
\end{table}

\paragraph{Passage pair construction and qrel scoring.}
All passage pairs for qrel construction were enumerated and scored by Jaccard similarity over the 9 discriminative NER categories (all except \textit{Values} and \textit{Conditions}, which provide weak compositional signal). Pairs with Jaccard $J \in [0.30, 0.50]$ were designated \emph{hard negatives} (score 2); pairs with $J \in [0.05, 0.15]$ were designated \emph{soft negatives} (score 1). The anchor passage (score 3) is the passage the query was generated from; score-0 passages are implicit (not written to qrel files, per BEIR convention). The cross-patent filter is enforced at pair-construction time: no pair from the same patent receives a non-zero score.

\subsection{Query Generation Pipeline}
\label{app:querygen}

\paragraph{Generation stage.}
Each candidate anchor passage was submitted to Claude Sonnet 4~\citep{anthropic2025sonnet4} via a two-step prompt. Step 1 (\emph{pre-analysis}): the model identifies the passage's formulation type (pharmaceutical, adhesive, coating, etc.), key ingredients, and primary technical function. Step 2 (\emph{query generation}): the model generates a retrieval question in one of four modes---\textit{composition} (``What components are used in\ldots''), \textit{process} (``How is\ldots prepared?''), \textit{performance} (``What are the mechanical/physical properties of\ldots''), or \textit{application} (``For what applications is\ldots suitable?''). The mode was sampled uniformly. The full generation prompt instructed the model to produce a question a formulation scientist or patent examiner would plausibly ask, avoiding direct quotes from the passage and avoiding questions whose answers require reading the full patent (rather than just the target passage).

\paragraph{Quality filtering.}
Candidate queries were filtered by Claude Haiku 4.5~\citep{anthropic2025haiku45} scoring each on three criteria, each on a binary pass/fail scale: (1) \emph{grammaticality} (the question is well-formed English with no truncation or incoherence); (2) \emph{specificity} (the question is specific to the passage's formulation content, not a generic chemistry question); (3) \emph{passage-match} (retrieving the source passage would plausibly answer the question). Queries failing any criterion were discarded. The filter pass rate was approximately 72\% across all passage types, with lower pass rates on process queries (66\%) where generated questions sometimes referred to fabrication details not present in the source passage.

\paragraph{Cost and throughput.}
The full query generation pipeline (generation + filtering) processed 77,000 candidate passages at a total API cost of \$904.41 USD, yielding 55,352 retained queries (72\% pass rate). The filtering pass adds approximately 15\% overhead to the generation cost.

\paragraph{Patent-level splitting.}
Queries were assigned to train/dev/test splits at the \emph{patent} level using a stratified 80/10/10 split by \texttt{clustered\_category} (the fine-grained taxonomy label of the anchor passage). All queries derived from the same patent land in exactly one split; no passage from a training patent can appear as a retrieval target in dev or test. Split disjointness was verified programmatically via set-intersection assertions on train, dev, and test patent sets (all three pairwise intersections were confirmed empty). This procedure yielded 44,413 / 5,480 / 5,459 queries in train / dev / test respectively.

\paragraph{Prompts.}
The full prompts used for pre-analysis, query generation, and two-stage quality filtering are provided in the project code repository (Appendix~\ref{app:maintenance}).

\section{Experimental Details}
\label{app:experiments}

\subsection{Model Architectures}

\begin{table}[h]
  \centering
  \small
  \caption{Open-source embedding models evaluated on \fb{}. All models use max sequence length 8,192; encoding uses 2,048 (covers 99.87\% of corpus passages without truncation).}
  \begin{tabular}{lllll}
    \toprule
    \textbf{Model} & \textbf{Params} & \textbf{Arch.} & \textbf{Pooling} & \textbf{Emb.\ dim} \\
    \midrule
    Nomic-embed-text-v1.5  & 137M & Encoder & Mean           & 768   \\
    GTE-large-en-v1.5      & 434M & Encoder & CLS            & 1,024 \\
    Qwen3-Embedding-0.6B   & 600M & Decoder & Last-token EOS & 1,024 \\
    \bottomrule
  \end{tabular}
  \label{tab:model-arch}
\end{table}

\noindent\small\textbf{HuggingFace model IDs:} Nomic: \texttt{nomic-ai/nomic-embed-text-v1.5};\quad GTE-large: \texttt{Alibaba-NLP/gte-large-en-v1.5};\quad Qwen3-Embed: \texttt{Qwen/Qwen3-Embedding-0.6B}.\normalsize

\subsection{Evaluation Protocol}

Open-source models encode the full corpus on OLCF Frontier (AMD MI250X GPUs) with batch size 32, max sequence length 2,048, each GPU rank encoding its own corpus shard independently (zero inter-rank communication during encoding). Commercial APIs encode via provider inference endpoints; latency reflects round-trip network + inference time and is not directly comparable to on-device GPU throughput.

Retrieval uses FAISS \texttt{IndexFlatIP} (exact inner-product search)~\citep{johnson2019faiss}, top-$k$\,=\,100, with 8 OMP threads on the OLCF DTN SSH node (376\,GB RAM). Metrics are computed with \texttt{pytrec\_eval} against test-split qrels (5,459 queries).

\subsection{Training Dynamics}
\label{app:training}
\paragraph{Note on DAPT.}
Prior to the contrastive TAPT runs reported in the main paper, we also explored domain-adaptive pretraining (DAPT) at two corpus scales, denoted D1 ($\sim$50M tokens) and D2 ($\sim$800M tokens). \textbf{D1} applied continued pretraining (MLM for Nomic and GTE; causal LM for Qwen3) on the $\sim$50M-token triplet corpus used for TAPT; \textbf{D2} applied the same objective on the full $\sim$800M-token C1 corpus. Neither scale produced retrieval gains over the zero-shot baselines: D1 was flat-to-slightly-negative across architectures, and D2 caused catastrophic forgetting (Nomic: binary nDCG@10 $0.1465 \!\to\! 0.0082$). A subsequent DAPT\,$\to$\,TAPT stage recovered most of the lost retrieval signal but did not exceed TAPT-only performance. Because DAPT offered no benefit on \fb{}, it is omitted from the main paper; we include the per-checkpoint trajectories here for completeness and as context for readers familiar with the DAPT+TAPT literature.
\subsubsection{DAPT Checkpoint Trajectories}

Per-checkpoint binary nDCG@10 scores for all DAPT runs are reported in Table~\ref{tab:dapt-checkpoints}.

\begin{table}[H]
  \centering
  \small
  \caption{Binary nDCG@10 (C1 test set) at each DAPT checkpoint. D1 = triplet corpus ($\sim$50M tokens); D2 = full C1 corpus ($\sim$800M tokens). Zero-shot baselines shown for reference. D2 for GTE and Qwen3 was not run after D2 Nomic showed a consistent degradation pattern.}
  \begin{tabular}{llrr}
    \toprule
    \textbf{Model} & \textbf{Checkpoint} & \textbf{bin-nDCG@10} & \textbf{vs.\ zero-shot} \\
    \midrule
    Nomic (zero-shot)      & ---        & 0.1465 & --- \\
    Nomic D1 ckpt-10       & 10\%       & 0.1467  & $+$0.0002 \\
    Nomic D1 ckpt-20       & 20\%       & 0.1470  & $+$0.0005 \\
    Nomic D1 ckpt-30       & 30\%       & 0.1474  & $+$0.0009 \\
    Nomic D1 ckpt-40       & 40\%       & 0.1472  & $+$0.0007 \\
    Nomic D1 final         & 100\%      & 0.1472  & $+$0.0007 \\
    \midrule
    Nomic D2 ckpt-193      & 25\%       & 0.1218  & $-$0.0247 \\
    Nomic D2 ckpt-386      & 50\%       & 0.0183  & $-$0.1282 \\
    Nomic D2 ckpt-579      & 75\%       & 0.0089  & $-$0.1376 \\
    Nomic D2 final         & 100\%      & 0.0082  & $-$0.1383 \\
    \midrule
    GTE (zero-shot)        & ---        & 0.2052 & --- \\
    GTE D1 ckpt-21         & 25\%       & 0.1807  & $-$0.0245 \\
    GTE D1 ckpt-42         & 50\%       & 0.1804  & $-$0.0248 \\
    GTE D1 ckpt-63         & 75\%       & 0.1814  & $-$0.0238 \\
    GTE D1 ckpt-84         & 88\%       & 0.1818  & $-$0.0234 \\
    GTE D1 final           & 100\%      & 0.1818  & $-$0.0234 \\
    \bottomrule
  \end{tabular}
  \label{tab:dapt-checkpoints}
\end{table}

The Nomic D2 trajectory reveals catastrophic forgetting: retrieval performance collapses from 0.1465 (zero-shot) to 0.0082 (D2 final) as the model absorbs 800M tokens of MLM signal, losing its discriminative geometry entirely. Despite this collapse, subsequent TAPT-MNRL fine-tuning recovers nearly all retrieval capability (D2$\to$TAPT: 0.3665 vs.\ TAPT-only: 0.3668), demonstrating that the embedding geometry can be rebuilt from a collapsed state by contrastive training. GTE D1 shows a milder but monotonic degradation throughout D1 training (0.1807 $\to$ 0.1818 across checkpoints, consistently below zero-shot 0.2052).

\subsubsection{TAPT Learning Curves}

\begin{figure}[H]
  \centering
  \includegraphics[width=0.72\linewidth]{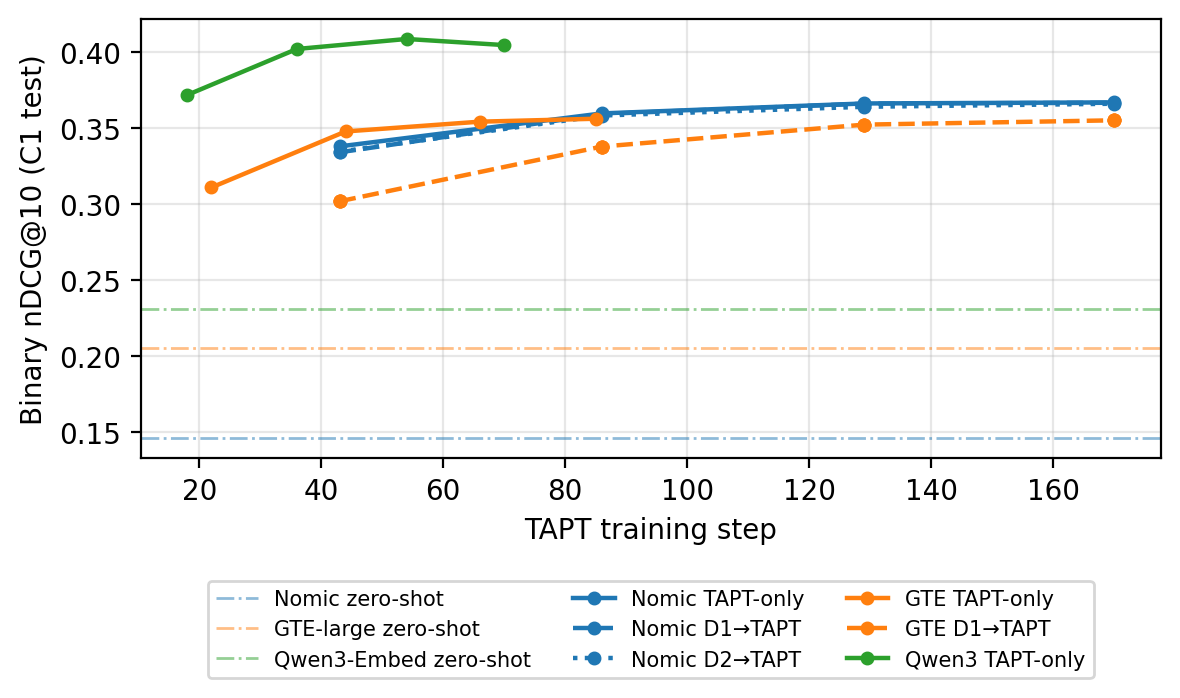}
  \caption{Binary nDCG@10 on \fb{} C1 (test set) across TAPT training steps for all adaptation configurations. DAPT-initialised models (dashed) converge below their TAPT-only counterparts (solid) for Nomic and GTE, confirming that DAPT initialisation hinders rather than helps contrastive fine-tuning. Dash-dot lines show zero-shot baselines for each architecture.}
  \label{fig:dapt-tapt}
\end{figure}

\begin{figure}[H]
  \centering
  \includegraphics[width=0.72\linewidth]{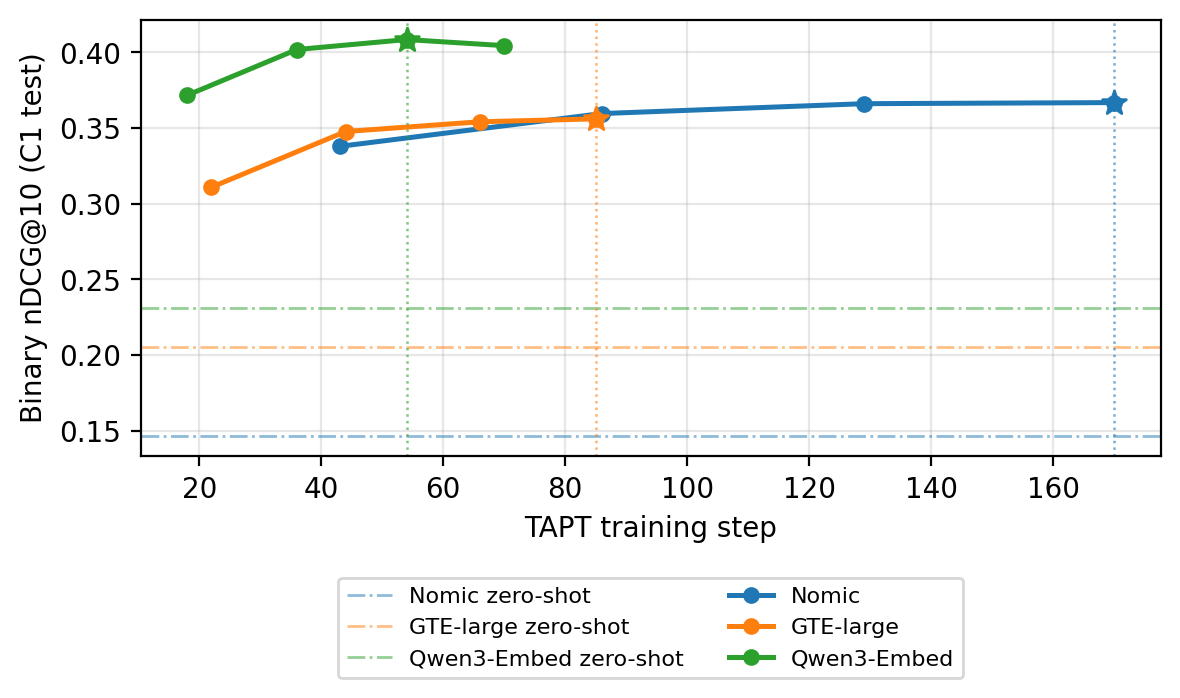}
  \caption{TAPT-MNRL learning curves on \fb{} C1 (test set). All three architectures converge within the first 50--60\% of training steps; Qwen3 peaks at ckpt-54 (75\% of steps) before modest decline, while Nomic and GTE continue improving to the full training budget. Stars mark best checkpoints; dash-dot lines show zero-shot baselines.}
  \label{fig:tapt-curves}
\end{figure}

Figures~\ref{fig:dapt-tapt} and~\ref{fig:tapt-curves} show per-checkpoint binary nDCG@10 on the C1 test set across TAPT training. Checkpoint numbering reflects effective update steps at the reported effective batch size; each model saved 4 evenly-spaced checkpoints over 5 training epochs.

\subsubsection{Hyperparameters}

DAPT and TAPT training hyperparameters are listed in Tables~\ref{tab:dapt-hparams} and~\ref{tab:tapt-hparams} respectively.

\begin{table}[H]
  \centering
  \small
  \caption{DAPT hyperparameters. Encoder models use masked language modelling (MLM, 15\% mask probability); Qwen3 uses causal LM. All runs on OLCF Frontier (AMD MI250X), DDP + DeepSpeed ZeRO-2, 3 epochs, bf16 precision.}
  \begin{tabular}{lrrr}
    \toprule
    & \textbf{Nomic} & \textbf{GTE-large} & \textbf{Qwen3-Embed} \\
    \midrule
    Learning rate           & 5$\times$10$^{-6}$ & 3$\times$10$^{-6}$ & 2$\times$10$^{-6}$ \\
    Batch size / GPU        & 32  & 64  & 32  \\
    Warmup ratio            & 5\% & 5\% & 5\% \\
    Sequence length         & 2,048 & 2,048 & 2,048 \\
    Precision               & bf16 & bf16 & bf16 \\
    Optimizer               & \multicolumn{3}{c}{AdamW (HF Trainer default)} \\
    Gradient checkpointing  & Yes & Yes & Yes \\
    Training framework      & \multicolumn{3}{c}{HF Trainer + DeepSpeed ZeRO-2} \\
    Loss objective          & MLM (15\% mask) & MLM (15\% mask) & Causal LM \\
    Checkpoints saved       & 4   & 4   & 4   \\
    Nodes (D1)              & 5   & 10  & 15  \\
    \bottomrule
  \end{tabular}
  \label{tab:dapt-hparams}
\end{table}

\begin{table}[H]
  \centering
  \small
  \caption{TAPT hyperparameters. All runs on OLCF Frontier, 5 epochs, bf16 precision. MNRL = MultipleNegativesRankingLoss~\citep{henderson2017mnrl} with in-batch negatives; Matryoshka~\citep{kusupati2022matryoshka} wraps MNRL with multi-resolution dimension training.}
  \begin{tabular}{lrrr}
    \toprule
    & \textbf{Nomic} & \textbf{GTE-large} & \textbf{Qwen3-Embed} \\
    \midrule
    Learning rate           & 2$\times$10$^{-5}$ & 1$\times$10$^{-5}$ & 1$\times$10$^{-5}$ \\
    Batch size / GPU        & 32  & 64  & 32  \\
    Warmup ratio            & 10\% & 10\% & 10\% \\
    Sequence length         & 2,048 & 2,048 & 2,048 \\
    Precision               & bf16 & bf16 & bf16 \\
    Optimizer               & \multicolumn{3}{c}{AdamW (HF Trainer default)} \\
    Gradient checkpointing  & Yes & Yes & Yes \\
    Training framework      & \multicolumn{3}{c}{SentenceTransformerTrainer} \\
    Training queries        & \multicolumn{3}{c}{44,413} \\
    Loss (MNRL)             & \multicolumn{3}{c}{MultipleNegativesRankingLoss, in-batch negatives} \\
    Matryoshka dims         & 768, 512, 256, 128, 64 & 1024, 512, 256 & 1024, 512, 256 \\
    Checkpoints saved       & 4   & 4   & 4   \\
    Nodes                   & 2   & 10  & 12  \\
    Best checkpoint (MNRL)  & ckpt-170 & ckpt-85 & ckpt-54 \\
    \bottomrule
  \end{tabular}
  \label{tab:tapt-hparams}
\end{table}

\clearpage

\section{Analysis Details}
\label{app:analysis}

\begin{figure}[H]
  \centering
  \includegraphics[width=0.72\linewidth]{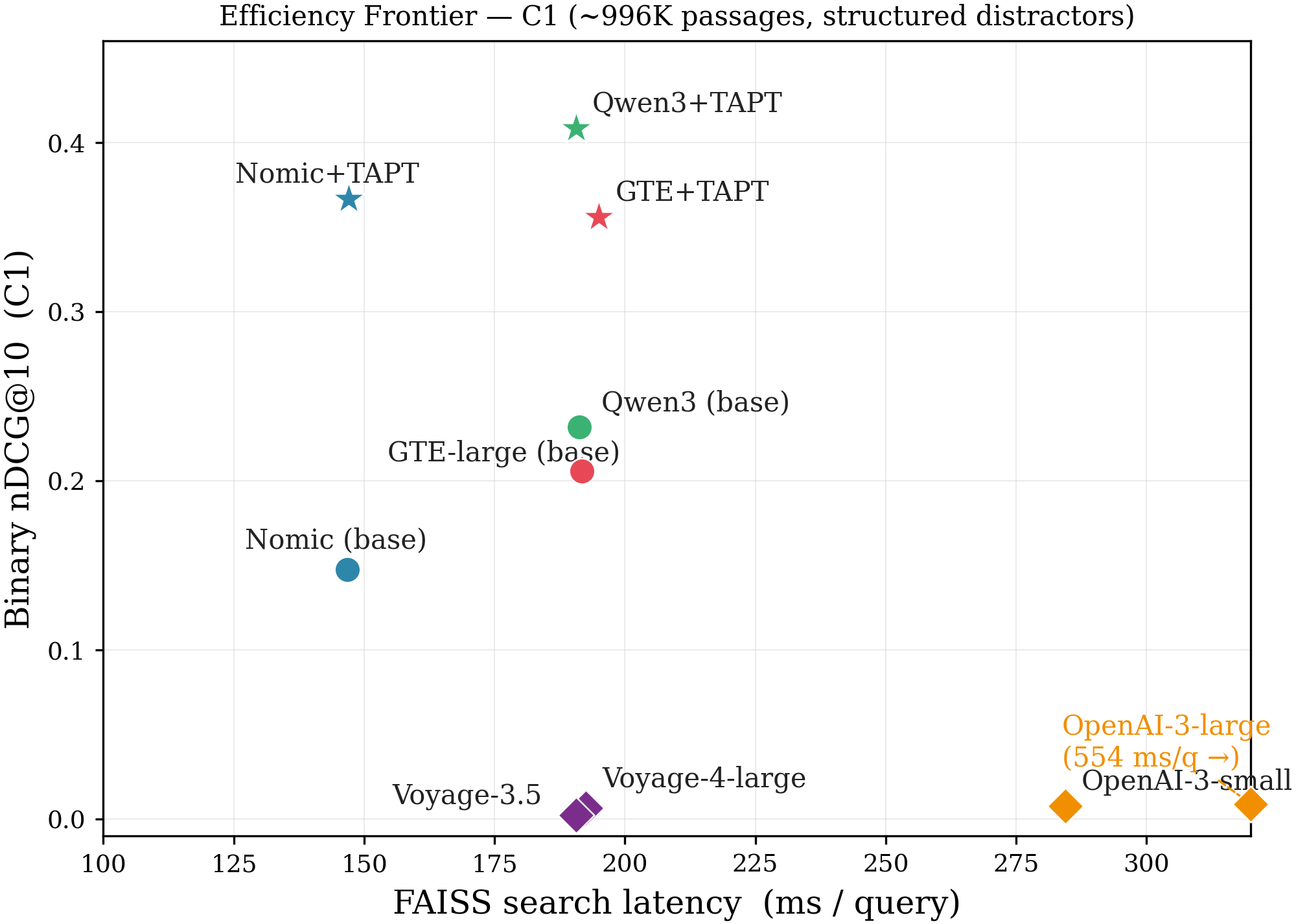}
  \caption{Efficiency frontier (C1): FAISS search latency vs.\ binary \ndcg{}. Open = base; filled = TAPT; diamond = API. TAPT models dominate the Pareto frontier.}
  \label{fig:efficiency}
\end{figure}

The efficiency frontier plot (latency vs.\ retrieval quality) is shown in Figure~\ref{fig:efficiency}. UMAP projections of corpus embeddings are shown in Figure~\ref{fig:umap}.

\subsection{CPC Clustering Results}
\label{app:clustering}

\begin{figure}[H]
  \centering
  \includegraphics[width=0.72\linewidth]{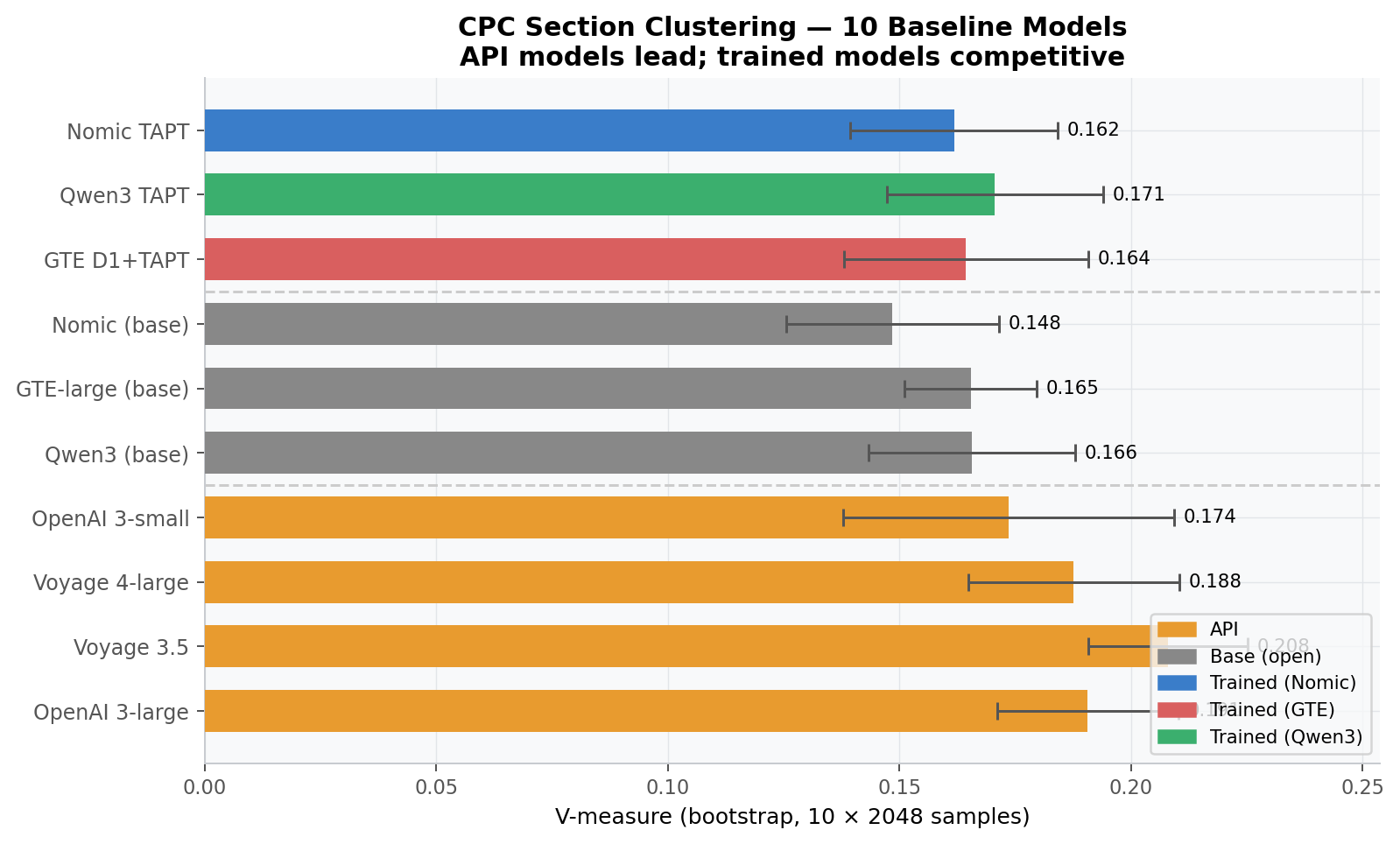}
  \caption{CPC section clustering (7 classes, bootstrapped V-measure, $n=10\times2{,}048$ samples). API models lead; TAPT adds no clustering advantage over zero-shot base models.}
  \label{fig:clu-section}
\end{figure}

\begin{figure}[H]
  \centering
  \includegraphics[width=0.72\linewidth]{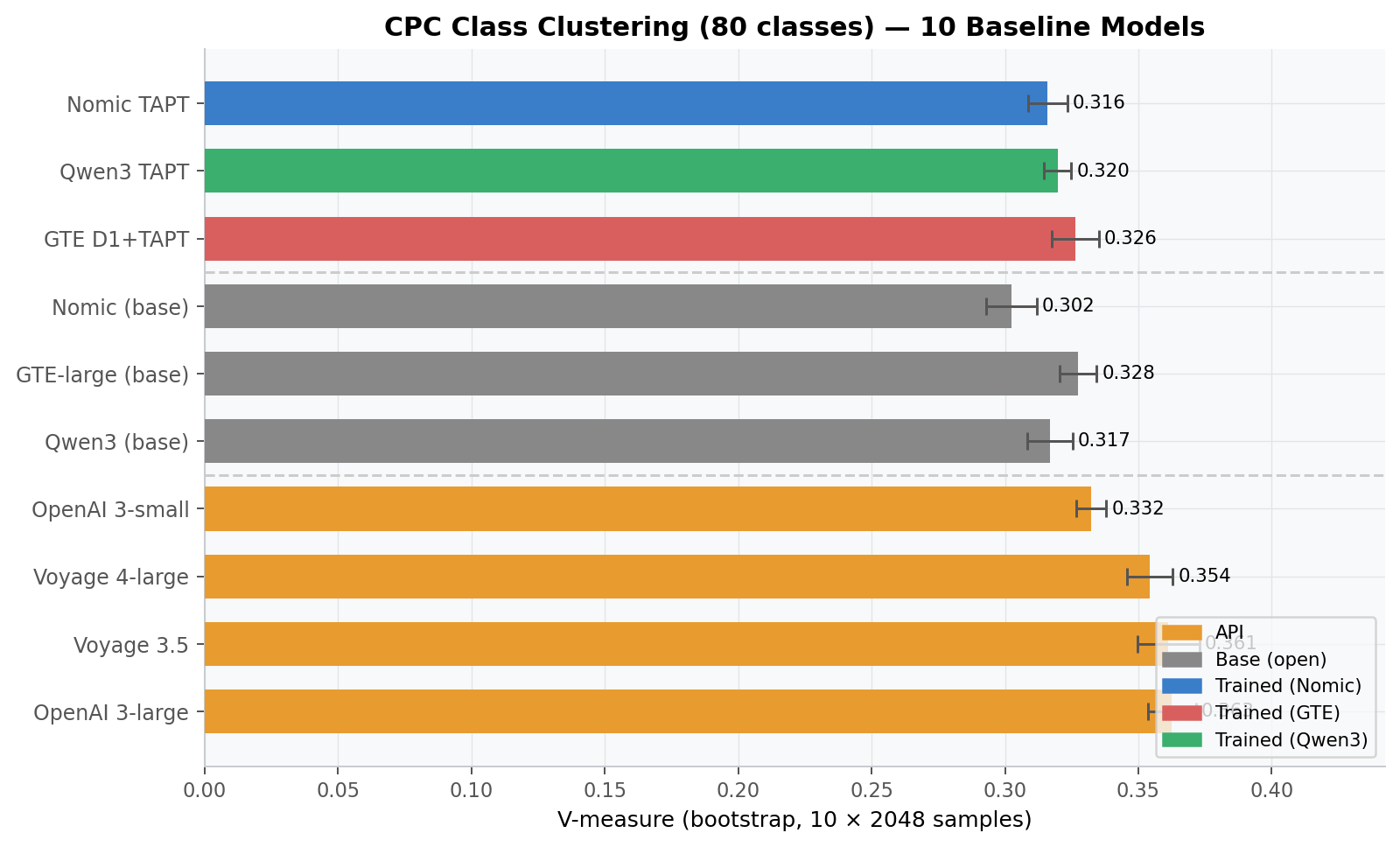}
  \caption{CPC subclass clustering (80 classes, V-measure). The API-leads pattern is stronger at finer granularity; TAPT models cluster no better than their zero-shot counterparts.}
  \label{fig:clu-class}
\end{figure}

\begin{figure}[H]
  \centering
  \includegraphics[width=0.72\linewidth]{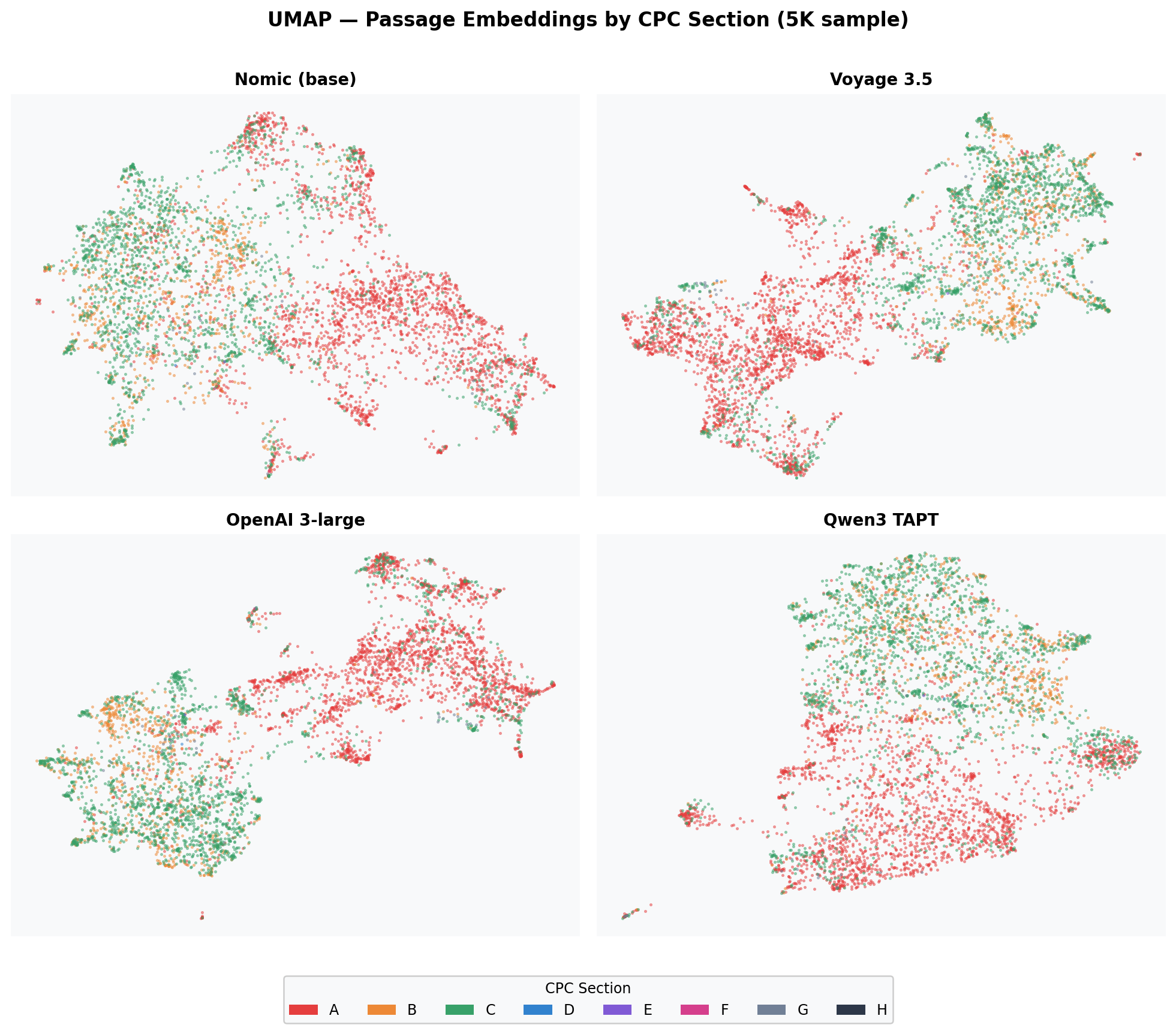}
  \caption{UMAP projections of C1 corpus passage embeddings for four representative models (zero-shot Qwen3-Embed, zero-shot Nomic, Qwen3-Embed TAPT-MNRL, Nomic TAPT-MNRL). Points coloured by CPC section. Zero-shot models show clear CPC-level clustering but no within-cluster separation; TAPT models develop finer-grained structure within each domain cluster, consistent with the retrieval--classification dissociation in Section~\ref{sec:downstream}.}
  \label{fig:umap}
\end{figure}

CPC clustering results are shown in Figures~\ref{fig:clu-section} and~\ref{fig:clu-class}. UMAP projections are shown in Figure~\ref{fig:umap}.

\section{Qualitative Retrieval Examples}
\label{app:qualitative}

Five case studies drawn from the C1 (near-miss distractor) test split, illustrating the behaviour of commercial API, open-source base, and TAPT-fine-tuned models on representative queries. The full set of 25 annotated case studies across all seven main-analysis models is released alongside the dataset and code (\texttt{results/qualitative\_case\_studies.md}). For each case below we report: the query; the anchor passage and the excerpt the query targets; anchor-rank for all seven models; and the top-1 retrieved passage (with a one-sentence analysis) for three representative models: OpenAI \texttt{text-embedding-3-large}, Nomic Embed base (zero-shot), and Qwen3-Embed TAPT-MNRL.

\subsection{Case \texttt{t7795\_B} --- architectural coatings}

\paragraph{Query.} \emph{What hydrophobic compounds are entrapped in wax seeded latex polymers to improve scuff resistance in architectural coatings?}

\paragraph{Domain.} Coatings \& Paints $\mid$ Coatings, Inks \& Adhesives.

\paragraph{Anchor.} \texttt{US11230645B2:0} --- \emph{Scuff resistant and chip resistant architectural compositions} (2022, CPC: C09D).
\begin{quote}\small
The present inventors have discovered that wax when used in certain or predetermined amounts and in combination with a hydrophobic aqueous latex resin produces paint films that resist scuffing and chipping. A preferred hydrophobic aqueous latex resin is disclosed in commonly owned U.S.\ Pat.\ No.\ 8,980,995 entitled ``Paints with Improved Water Staining and Color Rub-Off Qualities'' to Yang et al.\ldots{} The `995 patent teaches a hybrid latex polymer, wherein the latex comprises a hydrophobic or wax center entrapped by a polymeric shell, hereinafter ``wax seeded latex''\ldots{} The hydrophobic compounds can be silicones, silicone modified alkyds, low molecular weight polyethylene, low molecular weight polypropylene and paraffins\ldots{}
\end{quote}
\noindent\emph{Relevant excerpt:} ``The hydrophobic compounds can be silicones, silicone modified alkyds, low molecular weight polyethylene, low molecular weight polypropylene and paraffins, and the matrix polymer can be formed from any of the following monomers: (meth)acrylates, styrenated monomers, vinyl esters, and other ethylenically unsaturated monomers.''

\begin{center}\small
\begin{tabular}{lr}
\toprule
\textbf{Model} & \textbf{Anchor rank} \\
\midrule
OpenAI \texttt{text-embedding-3-large} & $>$100 \\
GTE-large (base)                       & 1 \\
Nomic (base)                           & 1 \\
Qwen3-Embed (base)                     & 1 \\
GTE-large (TAPT-MNRL)                  & 1 \\
Nomic (TAPT-MNRL)                      & 1 \\
Qwen3-Embed (TAPT-MNRL)                & 1 \\
\bottomrule
\end{tabular}
\end{center}

\paragraph{OpenAI \texttt{text-embedding-3-large} --- rank $>$100.}
Top-1: \texttt{US11230654B2:10} --- \emph{Calcium carbonate coated materials and methods of making and using same} (2022, CPC: A47L$\mid$C09K$\mid$C11D).
\begin{quote}\small
the WSF. Examples of oleaginous fluids suitable for use in a WSF include, but are not limited to, petroleum oils, natural oils, synthetically-derived oils\ldots{} paraffins, silicone fluids, polydiorganosiloxanes\ldots{}
\end{quote}
\emph{Analysis.} OpenAI latches onto surface-level chemistry vocabulary (paraffins, silicone fluids) that appears in a completely unrelated calcium carbonate coating context; the 2022 year and C09 classification proximity likely reinforce the false match, while the specific ``wax seeded latex'' / ``scuff resistance'' formulation concept is not recognised.

\paragraph{Nomic (base) --- rank 1.}
Top-1: \texttt{US11230645B2:0} --- \emph{Scuff resistant and chip resistant architectural compositions} (2022, CPC: C09D).
\begin{quote}\small
The present inventors have discovered that wax when used in certain or predetermined amounts\ldots{}
\end{quote}
\emph{Analysis.} Correct retrieval. The query's compound specificity (``wax seeded latex,'' ``entrapped,'' ``scuff resistance,'' ``architectural coatings'') is sufficient for Nomic-base to land the anchor at rank 1.

\paragraph{Qwen3-Embed (TAPT-MNRL) --- rank 1.}
Top-1: \texttt{US11230645B2:0} --- \emph{Scuff resistant and chip resistant architectural compositions} (2022, CPC: C09D).
\begin{quote}\small
The present inventors have discovered that wax when used in certain or predetermined amounts\ldots{}
\end{quote}
\emph{Analysis.} Correct retrieval. A $6/7$ success rate with OpenAI as the sole failure highlights that commercial general-purpose embeddings can underperform open-source models on highly domain-specific formulation queries.

\subsection{Case \texttt{t14820\_B} --- infrared-sensitive printing}

\paragraph{Query.} \emph{What electron migration mechanism causes color development in N,N-bis(p-aminoaryl)amide infrared-sensitive printing compositions?}

\paragraph{Domain.} Other $\mid$ Other.

\paragraph{Anchor.} \texttt{US10119028B2:0} --- \emph{Infrared-sensitive color developing composition, lithographic printing plate precursor, plate making method, and infrared-sensitive color developer} (2018, CPC: B41C$\mid$C07D$\mid$C08F$\mid$C09B$\mid$C09K).
\begin{quote}\small
\ldots{}the present inventors and the like assume that, since a structure having a conjugated chain in which $R_1$ to $R_8$ and the like bond to each other has a strong electron-accepting function, the compound represented by Formula~(1) is excited by exposure to infrared rays, electrons migrate from a structure having an N,N-bis(p-aminoaryl)amide structure to the structure having a conjugated chain\ldots{} and the N,N-bis(p-aminoaryl)amide structure is one-electron-oxidized in the molecule, whereby a decomposition reaction is caused, and a decomposed matter having a bis(p-aminoaryl)amine cation\ldots{} develops colors\ldots{}
\end{quote}
\noindent\emph{Relevant excerpt:} ``electrons migrate from a structure having an N,N-bis(p-aminoaryl)amide structure to the structure having a conjugated chain\ldots{} and the N,N-bis(p-aminoaryl)amide structure is one-electron-oxidized in the molecule, whereby a decomposition reaction is caused, and a decomposed matter having a bis(p-aminoaryl)amine cation\ldots{} develops colors.''

\begin{center}\small
\begin{tabular}{lr}
\toprule
\textbf{Model} & \textbf{Anchor rank} \\
\midrule
OpenAI \texttt{text-embedding-3-large} & $>$100 \\
GTE-large (base)                       & 10 \\
Nomic (base)                           & $>$100 \\
Qwen3-Embed (base)                     & 3 \\
GTE-large (TAPT-MNRL)                  & 1 \\
Nomic (TAPT-MNRL)                      & 1 \\
Qwen3-Embed (TAPT-MNRL)                & 1 \\
\bottomrule
\end{tabular}
\end{center}

\paragraph{OpenAI \texttt{text-embedding-3-large} --- rank $>$100.}
Top-1: \texttt{US10336886B2:1} --- \emph{Impact performance modified high transparent melt polymerized polycarbonate} (2019, CPC: C08G$\mid$C08K$\mid$F21V).
\begin{quote}\small
\ldots{}nducted. The salts of nonvolatile acids are alkali metal salts of phosphites; alkaline earth metal salts of phosphites\ldots{}
\end{quote}
\emph{Analysis.} A polycarbonate materials patent with no relation to infrared-sensitive printing. OpenAI likely fails because the highly specialised N,N-bis(p-aminoaryl)amide terminology and electron-migration mechanism concept have no clear counterpart in its training distribution.

\paragraph{Nomic (base) --- rank $>$100.}
Top-1: \texttt{US7563845B2:7} --- \emph{Amino resin crosslinked particles and producing process thereof} (2009, CPC: C08G$\mid$C08L).
\begin{quote}\small
---CH$_2$---NH--- bond (C(II) bond) to a carbon atom signal that originates from an ---NH---CH$_2$O---CH$_2$---NH--- bond\ldots{} a color difference in a heat discoloring test is not more than 15\ldots{}
\end{quote}
\emph{Analysis.} Nomic-base matches on ``amino'' and ``color'' surface features, retrieving an amino-resin discolouration-test passage; it misses the infrared / printing-plate / electron-migration specificity entirely and is drawn by generic amino-chemistry and colour-change language.

\paragraph{Qwen3-Embed (TAPT-MNRL) --- rank 1.}
Top-1: \texttt{US10119028B2:0} --- \emph{Infrared-sensitive color developing composition} (2018, CPC: B41C$\mid$C07D$\mid$C08F$\mid$C09B$\mid$C09K).
\begin{quote}\small
Hereinafter, the present invention will be described in detail\ldots{}
\end{quote}
\emph{Analysis.} Correct retrieval. All three TAPT-MNRL models achieve rank 1, up from a best of rank 3 among base models, showing that fine-tuning on domain triplets teaches the N,N-bis(p-aminoaryl)amide electron-mechanism concept as a retrieval signal.

\subsection{Case \texttt{t14087\_B} --- lubricant viscosity (hard case: all models fail)}

\paragraph{Query.} \emph{What maximum crystallite size is specified for molecular sieve catalysts used in low viscosity lubricant formulations?}

\paragraph{Domain.} Lubricants \& Oilfield $\mid$ Chemicals \& Energy.

\paragraph{Anchor.} \texttt{US7273834B2:6} --- \emph{Lubricant blends with low brookfield viscosities} (2007, CPC: C10M).
\begin{quote}\small
\ldots{}A particularly preferred intermediate pore size molecular sieve, which is useful in the present process is described, for example, in U.S.\ Pat.\ Nos.\ 5,135,638 and 5,282,958\ldots{} In U.S.\ Pat.\ No.\ 5,282,958, such an intermediate pore size molecular sieve has a crystallite size of no more than about 0.5 microns and pores with a minimum diameter of at least about 4.8\,\AA{} and with a maximum diameter of about 7.1\,\AA{}\ldots{}
\end{quote}
\noindent\emph{Relevant excerpt:} ``such an intermediate pore size molecular sieve has a crystallite size of no more than about 0.5 microns and pores with a minimum diameter of at least about 4.8\,\AA{} and with a maximum diameter of about 7.1\,\AA{}.''

\begin{center}\small
\begin{tabular}{lr}
\toprule
\textbf{Model} & \textbf{Anchor rank} \\
\midrule
OpenAI \texttt{text-embedding-3-large} & $>$100 \\
GTE-large (base)                       & $>$100 \\
Nomic (base)                           & 90 \\
Qwen3-Embed (base)                     & $>$100 \\
GTE-large (TAPT-MNRL)                  & $>$100 \\
Nomic (TAPT-MNRL)                      & $>$100 \\
Qwen3-Embed (TAPT-MNRL)                & $>$100 \\
\bottomrule
\end{tabular}
\end{center}

\paragraph{OpenAI \texttt{text-embedding-3-large} --- rank $>$100.}
Top-1: \texttt{US7772337B2:9} --- \emph{Method of formulating a molecular sieve catalyst composition by controlling component addition} (2010, CPC: B01J$\mid$C07C$\mid$C08F).
\begin{quote}\small
eferably from about 0.1\,$\mu$m to about 0.6\,$\mu$m. It is also desirable that the clay material have a particle size distribution such that $d_{90}$ can be less than about 1.5\,$\mu$m\ldots{}
\end{quote}
\emph{Analysis.} The passage discusses molecular sieve particle size in the context of FCC catalyst formulation, not lubricant viscosity; OpenAI correctly identifies the molecular-sieve crystallite-size subspace but misses the critical lubricant application context that distinguishes the anchor.

\paragraph{Nomic (base) --- rank 90.}
Top-1: \texttt{US10239051B2:2} --- \emph{Catalyst compositions comprising small size molecular sieves crystals deposited on a porous material} (2019, CPC: B01J$\mid$C07C$\mid$C10G).
\begin{quote}\small
was estimated at around 15\,nm to have at least one unit cell\ldots{} the minimum reachable size of a molecular sieve is in the range of 15\,nm to at most 50\,nm\ldots{}
\end{quote}
\emph{Analysis.} Nomic-base retrieves a molecular-sieve crystal-size passage focused on catalyst deposition, not lubricant formulation; the crystallite-size concept is matched but without the lubricant-viscosity application context, and the size regime differs (nm vs.\ 0.5\,$\mu$m).

\paragraph{Qwen3-Embed (TAPT-MNRL) --- rank $>$100.}
Top-1: \texttt{US10894246B2:0} --- \emph{Catalyst compositions comprising small size molecular sieves crystals deposited on a porous material} (2021, CPC: B01J$\mid$C07C$\mid$C10G).
\begin{quote}\small
As regards to the meaning of mesoporous, microporous and macroporous\ldots{} the term micropore refers to pore having a widths smaller than 2\,nm\ldots{}
\end{quote}
\emph{Analysis.} A molecular-sieve pore-classification passage with no lubricant context. All seven models fail on this query (every anchor rank $\geq$\,90), making it one of the hardest cases in the benchmark: the anchor's combination of molecular-sieve crystallite size with low-viscosity lubricant formulation is a rare cross-domain concept pair that C1 near-miss distractors obscure effectively.

\subsection{Case \texttt{q\_01891} --- automotive coolant corrosion}

\paragraph{Query.} \emph{What carboxylate chain length range provides corrosion protection for magnesium alloys in automotive coolant formulations?}

\paragraph{Domain.} Other $\mid$ Other.

\paragraph{Anchor.} \texttt{US12215272B2:1} --- \emph{Heat transfer fluids and corrosion inhibitor formulations for use thereof} (2025, CPC: C09K$\mid$C23F$\mid$H01M).
\begin{quote}\small
\ldots{}The corrosion inhibitor formulations typically used in conventional water/glycol-based coolants contain high concentrations of ionic species, such as silicates, nitrites, carboxylates (e.g., C$_4$--C$_{18}$ mono- or di-carboxylates, benzoates), molybdates, nitrates, phosphates, phosphonates, borates, and/or the like to provide corrosion protection for various metals in the cooling systems\ldots{}
\end{quote}
\noindent\emph{Relevant excerpt:} ``carboxylates (e.g., C$_4$--C$_{18}$ mono- or di-carboxylates, benzoates)\ldots{} to provide corrosion protection for various metals in the cooling systems.''

\begin{center}\small
\begin{tabular}{lr}
\toprule
\textbf{Model} & \textbf{Anchor rank} \\
\midrule
OpenAI \texttt{text-embedding-3-large} & $>$100 \\
GTE-large (base)                       & 1 \\
Nomic (base)                           & 1 \\
Qwen3-Embed (base)                     & 1 \\
GTE-large (TAPT-MNRL)                  & 1 \\
Nomic (TAPT-MNRL)                      & 1 \\
Qwen3-Embed (TAPT-MNRL)                & 2 \\
\bottomrule
\end{tabular}
\end{center}

\paragraph{OpenAI \texttt{text-embedding-3-large} --- rank $>$100.}
Top-1: \texttt{US12215273B2:0} --- \emph{High density brine internal phase} (2025, CPC: C09K$\mid$E21B).
\begin{quote}\small
Provided are compositions and methods for a well treatment and, more particularly, to oil-based treatment fluids that use a true crystallization temperature reduction additive\ldots{}
\end{quote}
\emph{Analysis.} A 2025 oilfield brine-treatment patent under the same C09K CPC classification and year as the anchor, leading OpenAI into a CPC/year similarity trap; the corrosion/magnesium/coolant chain is not recognised, while the surface similarity of industrial fluid formulation language triggers a false match.

\paragraph{Nomic (base) --- rank 1.}
Top-1: \texttt{US12215272B2:1} --- \emph{Heat transfer fluids and corrosion inhibitor formulations for use thereof} (2025, CPC: C09K$\mid$C23F$\mid$H01M).
\begin{quote}\small
\ldots{}magnesium alloys have the highest strength-to-weight ratio\ldots{}
\end{quote}
\emph{Analysis.} Correct retrieval.

\paragraph{Qwen3-Embed (TAPT-MNRL) --- rank 2.}
Top-1: \texttt{US11560505B2:26} --- \emph{Heat transfer fluids containing synergistic blends of corrosion inhibitor formulations} (2023, CPC: C09K$\mid$C23F).
\begin{quote}\small
EG 200, 300, 400, 600, 900, 1000, 1450, 3350, 4000 \& 8000\ldots{} Representative polyalkylene glycol esters include but are not limited to mono- and di-esters of various fatty acids\ldots{}
\end{quote}
\emph{Analysis.} A closely related heat-transfer-fluid corrosion-inhibitor patent displaces the anchor to rank 2. Qwen3+MNRL is the only configuration that regresses from the correct retrieval achieved by the base models, preferring the 2023 heat-transfer-fluid-blends patent over the 2025 magnesium-specific anchor --- plausibly because the 2023 patent family is more richly represented in the fine-tuning triplet distribution.

\subsection{Case \texttt{t15855\_A} --- phosgene production catalyst (same-patent confusion)}

\paragraph{Query.} \emph{What thermal conductivity range and coating thickness are specified for thermally conductive coatings on phosgene production catalyst particles?}

\paragraph{Domain.} Catalysts \& Chemical Processing $\mid$ Chemicals \& Energy.

\paragraph{Anchor.} \texttt{US10059594B2:3} --- \emph{Method and apparatus for producing high purity phosgene} (2018, CPC: B01J$\mid$C01B).
\begin{quote}\small
talyst and the tube. The thermally conductive material can have a thermal conductivity of greater than 2\,W/(m$\cdot$K), or greater than 15\,W/(m$\cdot$K), or greater than 100\,W/(m$\cdot$K), or greater than 200\,W/(m$\cdot$K). It can be a coating disposed on at least a portion of an exterior surface of the catalyst particles\ldots{} and can have a coating thickness of 0.001 to 1\,$\mu$m, and the coating can be deposited by chemical vapor deposition, thermal spraying, dip coating, or powder coating\ldots{}
\end{quote}
\noindent\emph{Relevant excerpt:} ``The thermally conductive material can have a thermal conductivity of greater than 2\,W/(m$\cdot$K)\ldots{} and can have a coating thickness of 0.001 to 1\,$\mu$m.''

\begin{center}\small
\begin{tabular}{lr}
\toprule
\textbf{Model} & \textbf{Anchor rank} \\
\midrule
OpenAI \texttt{text-embedding-3-large} & $>$100 \\
GTE-large (base)                       & 3 \\
Nomic (base)                           & 4 \\
Qwen3-Embed (base)                     & 3 \\
GTE-large (TAPT-MNRL)                  & 4 \\
Nomic (TAPT-MNRL)                      & 3 \\
Qwen3-Embed (TAPT-MNRL)                & 3 \\
\bottomrule
\end{tabular}
\end{center}

\paragraph{OpenAI \texttt{text-embedding-3-large} --- rank $>$100.}
Top-1: \texttt{US10059594B2:2} --- \emph{Method and apparatus for producing high purity phosgene} (2018, CPC: B01J$\mid$C01B).
\begin{quote}\small
aterial with the particulate catalyst. The thermally conductive particulate material and the particulate catalyst can be disposed in the reactor as a packed bed\ldots{}
\end{quote}
\emph{Analysis.} Correct patent, wrong passage: OpenAI retrieves chunk \texttt{:2} instead of chunk \texttt{:3} (the anchor). Both passages come from the same patent, but chunk \texttt{:2} describes the particulate-material arrangement without the specific thermal-conductivity range and coating thickness that the query targets.

\paragraph{Nomic (base) --- rank 4.}
Top-1: \texttt{US9598776B2:3} (patent metadata unavailable in source corpus).
\begin{quote}\small
[no text available]
\end{quote}
\emph{Analysis.} A passage for which patent metadata is missing occupies the top slot; the anchor at rank 4 indicates partial success in identifying the phosgene/catalyst/coating subspace.

\paragraph{Qwen3-Embed (TAPT-MNRL) --- rank 3.}
Top-1: \texttt{US10059594B2:2} --- \emph{Method and apparatus for producing high purity phosgene} (2018, CPC: B01J$\mid$C01B).
\begin{quote}\small
aterial with the particulate catalyst\ldots{}
\end{quote}
\emph{Analysis.} All non-OpenAI models identify the correct patent but fail to distinguish the specific coating-properties passage from the adjacent arrangement passage, suggesting that the quantitative thermal-conductivity and coating-thickness specification is not sufficiently differentiated from surrounding context in the current passage representations. This is a within-patent chunk-discrimination failure distinct from the cross-patent confusions exhibited by OpenAI.

\section{Validation of Graded Qrel Tiers}
\label{app:qrel-validation}

The graded qrel tier structure (anchor = 3, hard negative = 2, soft negative = 1) is derived from entity-Jaccard similarity between passage pairs. To verify that these tiers encode meaningful relevance gradations beyond their Jaccard construction, we run an LLM-as-judge pairwise-ranking study.

\paragraph{Protocol.}
We draw 1{,}000 (query, anchor, hard-negative, soft-negative) tuples from the test split, stratified across the six macro-domains ($\sim$167 tuples per domain; top-up from the remaining pool if a domain is short). For each tuple, the three passages are presented to GPT-4o (\texttt{gpt-4o-2024-08-06}) in one of the six possible orders, under a system prompt that instructs the model to rank the three passages by relevance to the query and provide a one-line justification. The judge returns structured JSON containing a 1-indexed ranking. Requests are submitted via the OpenAI Batch API at temperature 0; each passage is truncated to 4{,}500 characters to keep the context bounded. The total cost of the study was approximately USD 8.

We report three metrics per macro-domain and overall, with 1{,}000-sample bootstrap 95\% confidence intervals:
\begin{itemize}[leftmargin=*,itemsep=1pt,topsep=2pt]
  \item \textbf{A@1} --- fraction of tuples in which the judge placed the anchor first. Functions as a sanity check that the query-generation pipeline produces queries whose anchor is genuinely the most relevant passage.
  \item \textbf{HN\,$>$\,SN} --- fraction of tuples in which the judge placed the hard negative above the soft negative. This is the primary validation of the graded tier structure.
  \item \textbf{Perfect ordering} --- fraction of tuples in which the judge reproduced the full \mbox{A\,$>$\,HN\,$>$\,SN} ordering.
\end{itemize}

\paragraph{Results.}

Table~\ref{tab:qrel-validation} reports overall and per-macro-domain figures. Three observations. First, the judge places the anchor at rank~1 in 92.6\% of tuples overall [95\%\,CI 90.9--94.1]: the query-generation pipeline produces queries whose anchor is consistently the most relevant passage. Second, the judge places the hard negative above the soft negative in 64.7\% of tuples overall [61.8--67.8], above chance by a statistically meaningful margin (lower CI bound is 11.8\,pp above the 50\% chance line). Third, per-macro-domain HN\,$>$\,SN rates are tight (59.6--69.3\%) and their confidence intervals overlap; no domain shows a failure of the tier structure. 35\% of tuples are disagreements, which is expected given the tier boundary design: the HN tier spans Jaccard\,$\in [0.30, 0.50]$ and the SN tier Jaccard\,$\in [0.05, 0.15]$, so boundary cases (HN Jaccard near 0.30, SN Jaccard near 0.15) can be genuinely close in relevance. The LLM-as-judge study therefore validates that the graded tiers are meaningful on average but are not a precise proxy for human relevance judgements at fine granularity, supporting our decision to report binary \ndcg{} as the primary metric.

\begin{table}[h]
  \centering
  \small
  \caption{LLM-as-judge validation of graded qrel tiers on the test split, $n=1{,}000$ tuples stratified across six macro-domains; judge: GPT-4o (\texttt{gpt-4o-2024-08-06}), batch API, temperature 0. Values are percentages with bootstrap 95\% CIs in square brackets.}
  \label{tab:qrel-validation}
  \begin{tabular}{lrrr}
    \toprule
    \textbf{Macro-domain} & \textbf{$n$} & \textbf{A@1 (\%)} & \textbf{HN\,$>$\,SN (\%)} \\
    \midrule
    Chemicals \& Energy          & 166 & 91.0 [86.7,\,95.2] & 59.6 [52.4,\,66.9] \\
    Coatings, Inks \& Adhesives  & 166 & 93.4 [89.8,\,97.0] & 66.9 [60.2,\,74.7] \\
    Electronics \& Construction  & 166 & 92.2 [88.0,\,95.8] & 64.5 [56.6,\,71.7] \\
    Life Sciences \& Health      & 169 & 91.1 [86.4,\,95.3] & 65.7 [58.0,\,73.4] \\
    Materials \& Polymers        & 167 & 92.2 [88.0,\,96.4] & 62.3 [55.1,\,70.1] \\
    Other                        & 166 & 95.8 [92.2,\,98.8] & 69.3 [62.6,\,76.5] \\
    \midrule
    \textbf{Overall}             & \textbf{1{,}000} & \textbf{92.6 [90.9,\,94.1]} & \textbf{64.7 [61.8,\,67.8]} \\
    \bottomrule
  \end{tabular}
\end{table}

\noindent
Perfect ordering (A\,$>$\,HN\,$>$\,SN) is achieved in 60.3\% of tuples overall [57.6,\,63.8], consistent with the compound probability of A@1 and HN\,$>$\,SN. Justifications produced alongside each ranking are retained in the evaluation artefact released with the code repository and are available for per-case inspection.

\paragraph{Human validation on a 200-tuple sub-sample.}
To corroborate the LLM-as-judge findings with a non-LLM signal, a domain expert independently re-annotated a random 200-tuple sub-sample drawn from the same 1{,}000 (seed \texttt{SEED\_HUMAN=2024}; indices persisted in \texttt{human\_sample\_indices.json} for reproducibility). The annotator saw the same passage permutations as GPT-4o (paired comparison) and ranked the three passages from most to least relevant under blinded labels A/B/C, with no access to the tier identities, the GPT-4o ranking, or the model-generated justifications during annotation. Annotation was performed via the helper notebook \texttt{qrel\_validation\_human.ipynb} on the OLCF DTN over several sessions (approximately 10 hours total). The annotator is one of the paper's authors and a domain expert in materials chemistry.

Table~\ref{tab:qrel-validation-human} reports the human metrics together with GPT-4o restricted to the same 200-tuple sub-sample for direct paired comparison. Three observations. First, human A@1 is 78.6\% [72.3,\,83.7] and human HN\,$>$\,SN is 61.1\% [54.2,\,67.6], with the lower CI bound 4.2\,pp above the 50\% chance line; both rates remain above chance with margin under a domain-expert judge. Second, on the same 200 tuples GPT-4o reaches A@1 = 85.7\% [80.2,\,89.9] and HN\,$>$\,SN = 64.3\% [57.5,\,70.6]; the human and LLM intervals overlap on both metrics, indicating that humans and GPT-4o give comparable rates with humans being slightly more conservative on the top-1 call. Third, the paired human--GPT-4o agreement on the hard-vs-soft ordering --- the fraction of tuples where the two judges put HN and SN in the same relative order --- is 78.6\% [72.3,\,83.7], substantially above chance and the strongest single piece of evidence that the LLM-as-judge protocol is a reasonable proxy for domain-expert judgement at scale.

\begin{table}[h]
  \centering
  \small
  \caption{Human-expert validation on a 200-tuple sub-sample drawn from the same 1{,}000 used in Table~\ref{tab:qrel-validation} (seed \texttt{SEED\_HUMAN=2024}). GPT-4o numbers are recomputed on the same 200 tuples for paired comparison. Values are percentages with bootstrap 95\% CIs in square brackets.}
  \label{tab:qrel-validation-human}
  \begin{tabular}{lrrr}
    \toprule
    \textbf{Judge} & \textbf{$n$} & \textbf{A@1 (\%)} & \textbf{HN\,$>$\,SN (\%)} \\
    \midrule
    Human (domain expert)              & 200 & 78.6 [72.3,\,83.7] & 61.1 [54.2,\,67.6] \\
    GPT-4o (same 200-tuple sub-sample) & 200 & 85.7 [80.2,\,89.9] & 64.3 [57.5,\,70.6] \\
    \midrule
    \multicolumn{2}{l}{\textbf{Paired HN/SN ordering agreement}} & \multicolumn{2}{r}{\textbf{78.6 [72.3,\,83.7]}} \\
    \bottomrule
  \end{tabular}
\end{table}

\noindent
Perfect ordering under the human judge is 58.0\% [51.1,\,64.6], within 2.3\,pp of the 60.3\% obtained by GPT-4o on the full 1{,}000 and consistent with the compound probability of human A@1 and HN\,$>$\,SN.

\paragraph{Caveats.}
The human validation uses a single annotator. We do not report Cohen's $\kappa$ or other multi-annotator inter-rater agreement statistics; a study with a second domain expert annotating an overlapping sub-sample is the natural extension and is left for future work. Within this single-annotator constraint, the paired comparison with GPT-4o on identical tuples and identical permutations is the strongest evidence available: human and LLM-as-judge produce comparable validation rates on the same data, supporting the use of GPT-4o-at-scale as a reasonable proxy for the tier-validation question while acknowledging the precision limits of both judges at the fine-grained tier boundaries described above.

\section{Responsible AI}
\label{app:rai}

\paragraph{Data provenance and licensing.}
\fb{} is derived entirely from US patent documents in the USPTO public domain. US patent text is not subject to copyright protection under 17 U.S.C.~§105 and is freely redistributable. The dataset is released under CC-BY-4.0, which requires attribution but imposes no restrictions on commercial use or derivative works. The dataset does not include any data purchased from commercial data providers; all passages are derived from USPTO bulk download files available at \url{https://patentsview.org/download/data-download-tables}.

\paragraph{Personal information.}
Patent descriptions do not contain personally identifiable information (PII). Inventor names and assignee information appear in patent metadata but are not included in the corpus passages; \fb{} passages contain only the technical description text of each patent, which describes chemical formulations rather than individuals. No demographic data, health records, financial records, or other sensitive personal information is present.

\paragraph{Known biases and representational gaps.}
Four biases in \fb{} should be acknowledged: (1)~\emph{Geographic bias}: USPTO patents over-represent US-based assignees; European, Japanese, and South Korean formulation patent literature is not included, which may cause benchmark results to over-state performance for North American industrial formulation retrieval relative to global settings. (2)~\emph{Language bias}: all corpus passages and queries are English; multilingual formulation literature is entirely absent. (3)~\emph{Temporal bias}: the 1995--2022 window excludes the most recent formulation research; models trained on post-2022 data may show distribution shift effects not captured by the benchmark. (4)~\emph{Institutional bias}: large pharmaceutical companies and major chemical manufacturers with high patent output will be disproportionately represented; niche or academic formulation research that does not produce USPTO patents is absent.

\paragraph{Misuse potential.}
The benchmark itself does not constitute a hazardous resource. However, retrieval models fine-tuned on \fb{} could in principle be used to accelerate identification of chemically similar formulations at scale, which could raise competitive intelligence or intellectual property concerns in pharmaceutical or specialty chemical settings. The CC-BY-4.0 licence requires attribution but does not restrict commercial use; practitioners deploying \fb{}-fine-tuned models in proprietary contexts should conduct their own legal review.

\paragraph{Environmental impact.}
The computational work underlying \fb{} was performed on OLCF Frontier using a DOE allocation. GPU inference for the FAISS evaluation pipeline was also conducted on Frontier. Commercial API calls (OpenAI, Voyage, Claude) consumed approximately 1.2 billion tokens across all stages of construction and evaluation. The total carbon footprint has not been formally estimated; the OLCF allocation was funded under a DOE scientific computing programme and subject to OLCF's institutional carbon accounting.

\section{Maintenance Plan}
\label{app:maintenance}

\paragraph{Availability.}
The dataset, code repository and fine-tuned models described in this paper are currently private and will be made public upon publication of this work at a peer-reviewed venue. They will be released in two locations: the Albert Invent Hugging Face organisation (\texttt{albert-invent/FormBench} and \texttt{albert-invent/\{nomic,gte-large,qwen3-embed\}-formbench-mnrl}) and the authors' repositories (\texttt{vinven7/FormBench}, \texttt{vinven7/\{nomic,gte-large,qwen3-embed\}-formbench-mnrl}, \texttt{github.com/olivettigroup/formbench}).

\paragraph{Hosting and availability.}
\fb{} is hosted on HuggingFace under CC-BY-4.0 with indefinite hosting. The three corpus configurations (C0, C1, Sample) are available as separate dataset splits. All BEIR-format files (\texttt{corpus.jsonl}, \texttt{queries.jsonl}, \texttt{qrels/train.tsv}, \texttt{qrels/dev.tsv}, \texttt{qrels/test.tsv}) are downloadable via the HuggingFace \texttt{datasets} library with a single \texttt{load\_dataset} call.

\paragraph{Versioning policy.}
Releases are version-tagged (e.g., \texttt{v1.0}, \texttt{v1.1}) using HuggingFace dataset versioning. Within a major version, updates are backward-compatible: the query IDs, passage IDs, and qrel scores for existing records are not modified. New corpus passages or queries may be added in minor versions; any such additions are announced in the dataset card changelog. Breaking changes (e.g., revised qrel scoring criteria) require a major version increment. The current release is \texttt{v1.0}; downstream users should pin to a specific version tag for reproducible evaluation.

\paragraph{Evaluation tooling.}
A reference evaluation script is provided in the project code repository that reproduces all results in this paper from raw model embeddings. The script requires only \texttt{faiss-cpu}, \texttt{sentence-transformers}, and \texttt{pytrec\_eval} as dependencies and runs the full BEIR evaluation pipeline (encode, index, retrieve, score) for both C0 and C1 in a single invocation. Users wishing to evaluate new models on \fb{} need only implement the \texttt{encode(texts)} interface.

\paragraph{Trained models.}
The three TAPT-fine-tuned models reported in Table~\ref{tab:main-results} are released alongside the dataset under Apache 2.0 (inheriting the licence of their base models), one each for Nomic (137M), GTE-large (434M), and Qwen3-Embed (600M). Each is a sentence-transformers-format checkpoint loadable via \texttt{SentenceTransformer.from\_pretrained(\ldots)} and ships with a model card documenting training data, hyperparameters, and per-corpus evaluation results.

\paragraph{Contact.}
For dataset issues, bug reports, or version release enquiries, please contact the authors. Issues may also be filed via the HuggingFace dataset repository's Discussions tab.

\end{document}